\documentclass[12pt]{article}
\usepackage{bbm}
\usepackage{bbding}
\usepackage{amsmath}
\usepackage{amssymb}
\usepackage{amsfonts}
\usepackage{mathrsfs}
\usepackage{mathrsfs, amssymb}
\usepackage{booktabs}
\usepackage[perpage,symbol*]{footmisc}
\usepackage{graphicx}
\usepackage{multirow}
\usepackage[level]{datetime}
\usepackage[a4paper]{geometry}

\newcounter{theorem}
\newtheorem{theorem}{Theorem}[section]
\newcounter{lemma}

\newcounter{remark}
\newtheorem{remark}{Remark}[section]
\newcounter{example}

\newcounter{definition}

\newcounter{corollary}

\newcounter{proposition}
\newtheorem{proposition}{Proposition}[section]
\newcounter{assumption}

\newcounter{condition}

\newcounter{algorithm}

\makeatletter
\newcommand*{\indep}{%
  \mathbin{%
    \mathpalette{\@indep}{}%
  }%
}
\newcommand*{\nindep}{%
  \mathbin{
    \mathpalette{\@indep}{\not}
  }%
}
\newcommand*{\@indep}[2]{%
  \sbox0{$#1\perp\m@th$}
  \sbox2{$#1=$}
  \sbox4{$#1\vcenter{}$}
  \rlap{\copy0}
  \dimen@=\dimexpr\ht2-\ht4-.2pt\relax
  \kern\dimen@
  {#2}%
  \kern\dimen@
  \copy0 
}
\makeatother

\def\vtheta{\mbox{\boldmath$\theta$}}
\def\vvartheta{\mbox{\boldmath$\vartheta$}}

\def\b{\mbox{\boldmath$b$}}
\def\s{\mbox{\boldmath$s$}}

\def\V{\mbox{\boldmath$V$}}
\def\x{\mbox{\boldmath$x$}}
\def\y{\mbox{\boldmath$y$}}

\def\Y{\mbox{\boldmath$Y$}}

\def\var{\mbox{var}}

\def\prow{\stackrel{\textstyle p}{\longrightarrow}}
\def\asrow{\stackrel{a.s.}{\longrightarrow}}

\begin{document}
\baselineskip 18pt
\begin{center}
{\large\bf Semiparametric mean and variance joint models with clipped-Laplace link functions for bounded integer-valued time series}
\end{center}

\begin{center}
{ \textbf{\small{ Tianqing Liu}}$^*$}\\
{ \small{Center for Applied Statistical Research and School of Mathematics, Jilin University, China}}\\
{ \small{\texttt{\textsc{tianqingliu@gmail.com}}}}

{ \textbf{\small{ Xiaohui Yuan}}}\\
{ \small{School of Mathematics and Statistics, Changchun University of Technology, China}}\\
{ \small{\texttt{\textsc{yuanxh@ccut.edu.cn}}}}
\end{center}
\begin{center}
This version: \usdate\today
\end{center}

\footnotetext{$^*$Corresponding author, $^\dag$ equal authors contribution.}

\begin{abstract}
{\small We present a novel approach for modeling bounded count time series data, by deriving accurate upper and lower bounds for the variance of a bounded count random variable while maintaining a fixed mean. Leveraging these bounds, we propose semiparametric mean and variance joint (MVJ) models utilizing a clipped-Laplace link function. These models offer a flexible and feasible structure for both mean and variance, accommodating various scenarios of under-dispersion, equi-dispersion, or over-dispersion in bounded time series.
The proposed MVJ models feature a linear mean structure with positive regression coefficients summing to one and allow for negative regression cefficients and autocorrelations.  We demonstrate that the autocorrelation structure of MVJ models mirrors that of an autoregressive moving-average (ARMA) process, provided the proposed clipped-Laplace link functions with nonnegative regression coefficients summing to one are utilized. We establish conditions ensuring the stationarity and ergodicity properties of the MVJ process, along with demonstrating the consistency and asymptotic normality of the conditional least squares estimators. To aid model selection and diagnostics, we introduce two model selection criteria and apply two model diagnostics statistics. Finally, we conduct simulations and real data analyses to investigate the finite-sample properties of the proposed MVJ models, providing insights into their efficacy and applicability in practical scenarios.
\paragraph{\small Keywords:} Bounded integer-valued time series, Clipped-Laplace link functions, Conditional mean, Conditional variance, Semiparametric model.}
\end{abstract}

\section {Introduction} \label{sec1} \setcounter {equation}{0}
\def\theequation{\thesection.\arabic{equation}}

Time series data with bounded integer-valued observations are often encountered in various fields such as finance, healthcare, and ecology. Classical Gaussian regression models cannot capture the discreteness and boundedness of the data and thus are not suited for the modelling of such bounded count-valued series. Over the last few decades, researchers have developed models and methods that can properly account for the discreteness and boundedness of the data, see Wei${\ss}$ (2018) for a survey.

McKenzie (1985) first proposed a binomial thinning operator-based auto-regressive (BAR) model for analyzing bounded integer-valued time series. Wei${\ss}$ \& Pollett (2014) extended the BAR(1) model to incorporate state-dependent thinning probabilities, broadening its applicability. Risti\'{c} et al. (2016) introduced an integer-valued ARCH model characterized by a conditional binomial distribution, suitable for modeling counts with under-, equi-, or overdispersion. Kang et al. (2023) developed a BAR-hidden Markov model tailored for zero-and-one inflated bounded count time series. However, it's noted that achieving negative autocorrelation function (ACF) values with these models typically requires stringent parameter constraints to ensure non-negativity and boundedness in the data-generating process.

To relax such restrictions on the parameter and ACF values, a convenient way is to relax the linearity assumption on the conditional mean. By allowing for nonlinear assumption on the conditional mean, researchers can potentially capture more complex patterns in the data and improve the model's ability to handle negative autocorrelation and maintain non-negativity and boundedness. Chen et al. (2020) proposed binomial logit-ARCH models, which use a logit link to ensure the boundedness of the range. Wei${\ss}$ \& Jahn (2022) proposed a soft-clipping BGARCH model based on the soft-clipping function (Klimek \& Perelstein, 2020), which can produce both an approximately linear mean structure and negative ACF. By using the soft-clipping function,  Chen (2023) developed a discrete beta GARCH model that can model bounded time series with under-dispersion, equi-dispersion or over-dispersion. However, parameter values of this model are difficult to interpret. Since closed formulae for the conditional mean and  conditional variance are not available.

All the above models are constructed by modeling the conditional mean, and the conditional variance is a deterministic function of the conditional mean. It implies that once you know the conditional mean, you can determine the conditional variance precisely. The following two propositions account for this phenomenon.
\begin{proposition}\label{boundedcountprop}
(Bounded count properties) Let $X$ be a count random variable taking values in $\{0,1,\cdots,d\}$. Let $\mu=E(X)$ and $\sigma^2(\mu)=E(X-\mu)^2$ denote its mean and variance, respectively. By Markov's inequality, $\mathbb{P}(X=0|\mu)=1-\mathbb{P}(X>\epsilon|\mu)\geq 1-\min(1,\mu/\epsilon)$ and $\mathbb{P}(d-X=0|\mu)=1-\mathbb{P}(d-X>\epsilon|\mu)\geq 1-\min(1,(d-\mu)/\epsilon)$ for any $0<\epsilon<1$. Thus, $(\mathbbm{a})$: $\lim_{\mu\rightarrow0^+}\mathbb{P}(X=0|\mu)=\lim_{\mu\rightarrow d^-}\mathbb{P}(X=d|\mu)=1$ and $(\mathbbm{b})$: $\lim_{\mu\rightarrow0^+}\sigma^2(\mu)=\lim_{\mu\rightarrow d^-}\sigma^2(\mu)=0$ .
\end{proposition}
\begin{proposition}\label{boundedcountvar}
(The bounds of the variance) For $\mu\in(0,d)$, let $\mathcal {F}(\mu)$ be a collection of
integer-valued random variables such that $\mathcal {F}(\mu)=\{X:
E(X)=\mu, \mathbb{P}(X\in\{0,1,\cdots,d\})=1\}$.
Then,
\begin{eqnarray}\label{varbound}
R(\mu)\leq \var(X)\leq \mu(d-\mu),\ \text{for}\ X\in \mathcal {F}(\mu)\ \text{with}\ 0<\mu<d,
\end{eqnarray}
where
for $c\in \mathbb{R}$,
\begin{eqnarray}\label{Rfun}
R(c)=\{\Delta(c)+1-c\}\{c-\Delta(c)\}\ \text{with}\ \Delta(c)=\max\{z\in \mathbb{Z}: z\leq c\}.
\end{eqnarray}
Furthermore, $\lim_{\mu\rightarrow0^+}R(\mu)=\lim_{\mu\rightarrow d^-}R(\mu)=0$ and $\lim_{\mu\rightarrow0^+}\mu(d-\mu)=\lim_{\mu\rightarrow d^-}\mu(d-\mu)=0$.
\end{proposition}
Proposition \ref{boundedcountprop} states that, the variance of a bounded count random variable is constrained by its mean $\mu$, particularly when $\mu$ is close to the barriers 0 and $d$. Proposition \ref{boundedcountvar} implies that the variance $\sigma^2(\mu)$ is precisely bounded by two functions of $\mu$. In the following, we show that these upper and lower bounds of the variance can be achieved by considering the variances of two specific count random variables within the set $\mathcal {F}(\mu)$.

\begin{remark}\label{boundequa}
Let $U_0$ denote a uniform random variable with the support set $[0,1]$. Liu $\&$ Yuan (2013) proposed the following first-order random rounding operator:
\begin{eqnarray}
\odot_1(\mu,U_0)&=&\Delta(\mu)+1(U_0\geq 1+\Delta(\mu)-\mu),\ \mu\in
\mathbb{R},\label{round1}
\end{eqnarray}
where $1(A)$ is the indicator function of $A$. It is easy to verify that $\odot_1(\mu,U_0)\in\mathcal {F}(\mu)$ for $\mu\in(0,d)$ and it holds that $\var\{\odot_1(\mu,U_0)\}=R(\mu)$. Moreover, define
\begin{eqnarray}\label{binary}
\odot_{0,d}(\mu,U_0)=d1\bigr(U_0> 1-\frac{\mu}{d}\bigr),\ \mu\in(0,d).
\end{eqnarray}
Then, we get $\odot_{0,d}(\mu,U_0)\in \mathcal {F}(\mu)$ and it holds that $\var\{\odot_{0,d}(\mu,U_0)\}=\mu(d-\mu)$.
\end{remark}
From propositions \ref{boundedcountprop} and \ref{boundedcountvar}, it becomes evident that modeling the conditional mean and variance simultaneously for bounded integer-valued time series is challenging due to the strict restrictions outlined in (\ref{varbound}). To address this gap, we propose semiparametric mean and variance joint (MVJ) models with clipped-Laplace link functions for bounded counts time series. This approach offers flexibility in structuring both mean and variance while ensuring that the resulting conditional mean variance adheres to the restrictions stated in (\ref{varbound}). The incorporation of the proposed clipped-Laplace link functions facilitates achieving negative dependence for the MVJ model, thus enhancing its ability to capture complex relationships within the data.

The remaining of the paper is organized as follows. In Section 2, we first propose a mean and variance joint model for bounded integer-valued random variables and then proposed a clipped-Laplace link function for modeling their mean. Next, in Section 3, we propose the MVJ($p_1,p_2$) model for the bounded integer-valued time series. The stochastic properties of the proposed models are established and model selection methods for the MVJ($p_1,p_2$) models are discussed. Simulations are presented in Section 4. In Section 5,
We apply the proposed model to two real data sets. Finally, we conclude the paper in Section 6. The proofs of all
forthcoming results are postponed to the Section S1 of the Supplementary Material.

\section{A mean and variance joint model and the clipped-Laplace link} \setcounter {equation}{0}
\def\theequation{\thesection.\arabic{equation}}

In this section, we first propose a mean and variance joint model for count-valued observations
and then present the clipped-Laplace link functions for modeling their mean.

\subsection{A mean and variance joint model}

To motivate our MVJ model, let us start with a look at bounded integer-valued random variables. Let $\mathbb{Z}=\{\cdots,-1,0,1,\cdots\}$ denote the set of all integers and $a\in\mathbb{Z}$ and $b\in\mathbb{Z}$ be two constants such that $-\infty<a<b< +\infty$. The bounded random variable $Y\in\{a,\cdots,b\}$ can be written as
\begin{eqnarray}\label{srt}
Y=a+D ,
\end{eqnarray}
where $D$ is a count random variable with support set $\Omega_D=\{0,1,\cdots,d\}$ with $d=b-a$. Thus, we only need to model the bounded count variable $D$.

Let $\mu=E(D)$. For $0\leq \kappa_1\leq\mu< \kappa_2\leq d$, define
\begin{eqnarray}\label{binary}
X(\kappa_1,\kappa_2,U_0)=\kappa_1{1}\biggr(U_0\leq \frac{\kappa_2-\mu}{\kappa_2-\kappa_1}\biggr)+\kappa_2{1}\biggr(U_0> \frac{\kappa_2-\mu}{\kappa_2-\kappa_1}\biggr),
\end{eqnarray}
where $U_0$ is a uniform random variable with the support set $[0,1]$. Then, it is easy to verify that
\begin{eqnarray}
&&E\{X(\kappa_1,\kappa_2,U_0)\}=\mu\ \text{and}\ \var\{X(\kappa_1,\kappa_2,U_0)\}=(\mu-\kappa_1)(\kappa_2-\mu).\label{muvarx}
\end{eqnarray}
Obviously,
\begin{eqnarray*}\label{varboundconstruct}
R(\mu)\leq \var\{X(\kappa_1,\kappa_2)\}=(\mu-\kappa_1)(\kappa_2-\mu)\leq \mu(d-\mu),
\end{eqnarray*}
where $\var\{X(\kappa_1,\kappa_2)\}=R(\mu)$ if and only if $(\kappa_1,\kappa_2)=(\Delta(\mu),\Delta(\mu)+1)$ and $\var\{X(\kappa_1,\kappa_2)\}= \mu(d-\mu)$ if and only if $(\kappa_1,\kappa_2)=(0,d)$.

To generate patterns rich enough from sample paths, we define the random variable $r$ with its support set contained in the interval [0,1], which includes the special case where $r$ is a constant.
Based on $r$, define
\begin{eqnarray}\label{kfun}
\kappa_1(\mu,r,U_1)=\odot_1((1-r)\Delta(\mu),U_1)\ \text{and}\ \kappa_2(\mu,r,U_2)=\odot_1((1-r)\{\Delta(\mu)+1\}+r d,U_2),
\end{eqnarray}
where the first-order random rounding operator $\odot_1$ is given in (\ref{round1}); $r$, $U_1$ and $U_2$ are mutually independent; and $U_1$ and $U_2$ are uniform random variables with a common support set $[0,1]$.
By the definition of $\odot_1$, we get
\begin{eqnarray}\label{kfunbound}
\kappa_1(\mu,r,U_1)\in[0,\Delta(\mu)]\cap\mathbb{Z}\ \text{and}\ \kappa_2(\mu,r,U_2)\in[\Delta(\mu)+1,d]\cap\mathbb{Z}.
\end{eqnarray}
Define
\begin{eqnarray}
D(\mu,r,U_0,U_1,U_2)=X(\kappa_1(\mu,r,U_1),\kappa_2(\mu,r,U_2),U_0),\label{Dexpre}
\end{eqnarray}
where $U_s$, $s=0,1,2$, are independent uniform random variables defined previously.
From (\ref{muvarx}), the law of iterated expectations and the law of total variance, we obtain
\begin{eqnarray}
E\{D(\mu,r,U_0,U_1,U_2)|r\}=\mu\ \ \text{and}\ \ \var\{D(\mu,r,U_0,U_1,U_2)|r\}=\Psi_r(\mu|d),\label{muvarxr}
\end{eqnarray}
where
\begin{eqnarray}
\Psi_r(\mu|d)&=&\{\mu-(1-r)\Delta(\mu)\}[(1-r)\{\Delta(\mu)+1\}+r d-\mu].\label{Psimu}
\end{eqnarray}
Based on (\ref{muvarxr}), (\ref{Psimu}), using the law of iterated expectations and the law of total variance again, it follows that
\begin{eqnarray}
&&E\{D(\mu,r,U_0,U_1,U_2)\}=\mu,\ \ \text{and}\label{cmuxr}\\
&&\var\{D(\mu,r,U_0,U_1,U_2)\}=E\{\Psi_r(\mu|d)\}\nonumber\\
&&=E\bigr(\{\mu-\Delta(\mu)+r\Delta(\mu)\}\bigr[\Delta(\mu)+1-\mu+r\{d-\Delta(\mu)-1\}\bigr]\bigr)\nonumber\\
&&=R(\mu)+E(r^2)\Delta(\mu)\{d-\Delta(\mu)-1\}\nonumber\\
&&+E(r)\bigr[\{\mu-\Delta(\mu)\}\{d-\Delta(\mu)-1\}+\Delta(\mu)\{\Delta(\mu)+1-\mu\}\bigr],
\label{cvarxr}
\end{eqnarray}
where $\mu-\Delta(\mu)\geq0$, $\Delta(\mu)+1-\mu\geq0$, $\Delta(\mu)\geq0$ and $d-\Delta(\mu)-1\geq0$, for $\mu\in[0,d)$. Moreover, it is easy to verify that
\begin{eqnarray*}
&&\lim_{\mu\rightarrow0^+}\var\{D(\mu,r,U_0,U_1,U_2)|r\}=\lim_{\mu\rightarrow0^+}\Psi_r(\mu|d)=0,\\
&&\lim_{\mu\rightarrow d^-}\var\{D(\mu,r,U_0,U_1,U_2)|r\}=\lim_{\mu\rightarrow d^-}\Psi_r(\mu|d)=0,
\end{eqnarray*}
which implies that
\begin{eqnarray*}
&&\lim_{\mu\rightarrow0^+}\var\{D(\mu,r,U_0,U_1,U_2)\}=\lim_{\mu\rightarrow0^+}E\{\Psi_r(\mu|d)\}=0\\
&&\lim_{\mu\rightarrow d^-}\var\{D(\mu,r,U_0,U_1,U_2)\}=\lim_{\mu\rightarrow d^-}E\{\Psi_r(\mu|d)\}=0.
\end{eqnarray*}
In the following, we present our findings on the basic properties of the conditional variance function $\Psi_r$ defined in (\ref{Psimu}).
\begin{proposition}\label{varfun}
For fixed $\mu \in [0, d)$,
$\Psi_r(\mu|d)$ is a nondecreasing function of $r$ over the interval $[0,1]$. Moreover, $\Psi_0(\mu|d)=R(\mu)$ and $\Psi_{1}(\mu|d)=\mu(d-\mu)$. Thus, $\Psi_r(\mu|d)$ exhausts all conceivable patterns related to the mean and variance structure of a bounded count random variable. On the other hand, $r$ can be interpreted as a dispersion hyper-parameter that influences the variance of the random variable $D(\mu,r,U_0,U_1,U_2)$. In Figure 1, we present the plots of the function $\Psi_r(\cdot|3)$ with different values of $r$.
\end{proposition}

\subsection{The clipped-Laplace link}

Let $\{D_t\in\{0,1\cdots,d\}: t \in \mathbb{Z}\}$ be a count-valued time series  and $\mathcal {F}_{t-1}$ be the $\sigma$-field generated by
$\{D_{t-1}, D_{t-2},\cdots\}$. Let $\mathbb{R}$ denote the set of all real numbers. Suppose $E(D_t|\mathcal {F}_{t-1})=M(\xi_t)$, where $\xi_t\in\mathbb{R}$ is a $\mathcal {F}_{t-1}$ measurable function. Since $D_t\in\{0,1\cdots,d\}$, the link function $M(\cdot)$ should be a measurable function such that $M(\cdot): \mathbb{R}\mapsto [0,d]$.

A convenient choice for the link function $M(\cdot)$ is the clipped ReLU activation function (Cai et al., 2017), i.e., $\mathrm{CReLU}(u|d)=\min(\max(u,0),d)$ with $d\in \mathbb{Z}_+\setminus\{0\}$. The clipped softplus function $\mathrm{CS}_\sigma(u|d)=\sigma\log\bigr\{\frac{1+\exp(\frac{u}{\sigma})}{1+\exp(\frac{u-d}{\sigma})}\bigr\}$ with $u\in\mathbb{R}$ and $\sigma>0$ (Klimek \& Perelstein, 2020) is also applicable. Let $F(u)=0.5\exp(u){1}(u\leq0)+\{1-0.5\exp(-u)\}{1}(u>0)$ be the standard Laplace cumulative distribution function (CDF). Liu \& Yuan (2024) proposed the Laplace link function $\mathrm{L}_\sigma(u)=-\sigma\log\{1-F(u/\sigma)\}$ with adjustment parameter $\sigma> 0$ for modeling the means of unbounded count random variables. The Laplace link function is linear over $u\in[0,+\infty)$. However, the Laplace link cannot be applied to bounded counts as the Laplace link function is not bounded from above. Following Liu \& Yuan (2024), we propose the following clipped-Laplace link function:
\begin{eqnarray}\label{CLaplace}
\mathrm{CL}_\sigma(u|d)=s(\sigma|d)\{\mathrm{L}_\sigma(u)-u-\mathrm{L}_\sigma(d-u)\}+0.5d\{1+s(\sigma|d)\},\ u\in\mathbb{R},\ \ \sigma>0,
\end{eqnarray}
where
\begin{eqnarray}\label{Laplaceconstant}
s(\sigma|d)=\frac{0.5d}{0.5d+\sigma\log(2)},\ \ \sigma>0,
\end{eqnarray}
and $\sigma = 1$ is the default choice. $\mathrm{CL}_\sigma(\cdot|1)$ is a CDF and Figure 2 gives the plots of $\mathrm{CL}_\sigma(\cdot|1)$ with different values of $\sigma$. It is easy to verify that
\begin{eqnarray}\label{Clinear}
\mathrm{CL}_\sigma(u|d)=s(\sigma|d)u+0.5d\{1-s(\sigma|d)\},\ \text{for}\ u\in[0,d]\ \text{and}\ \sigma>0.
\end{eqnarray}
Thus, $\mathrm{CL}_\sigma(u|d)$ is linear over $u\in[0,d]$. Furthermore, $\mathrm{CL}_\sigma(u|d)$ is a continuously differentiable function with respect to $u$ and it holds that
\begin{eqnarray}\label{pClinear}
&&\mathrm{CLD}_\sigma(u|d)=:\frac{\partial\mathrm{CL}_\sigma(u|d)}{\partial u}\nonumber\\
&=&s(\sigma|d)\{\mathrm{L}_\sigma'(u){1}(u<0)+{1}(0\leq u\leq d)+\mathrm{L}_\sigma'(d-u){1}(u>d)\}\geq0,
\end{eqnarray}
where $u\in\mathbb{R}$, $\sigma>0$, $\mathrm{L}_\sigma'(u)=\partial\mathrm{L}_\sigma(u)/\partial u=\frac{f(u/\sigma)}{1-F(u/\sigma)}$ and $f(u)=0.5\exp(-|u|)$ is the standard Laplace density function. $\mathrm{CLD}_\sigma(\cdot|1)$ is a density function and  Figure 3 presents the plots of $\mathrm{CLD}_\sigma(\cdot|1)$ with different values of $\sigma$.

The following three equalities hold:
\begin{eqnarray}\label{limitbound}
\lim_{u\rightarrow-\infty}\mathrm{CL}_\sigma(u|d)=0,\ \lim_{u\rightarrow+\infty}\mathrm{CL}_\sigma(u|d)=d\ \text{and}\ \lim_{\sigma\rightarrow0^+}\mathrm{CL}_\sigma(u|d)=\mathrm{CReLU}(u|d).
\end{eqnarray}
In fact,
\begin{eqnarray*}
\lim_{u\rightarrow-\infty}\mathrm{CL}_\sigma(u|d)&=&s(\sigma|d)[-u-\{d-u+\sigma\log(2)\}]+0.5d\{1+s(\sigma|d)\}\\
&=&s(\sigma|d)\{-\sigma\log(2)-0.5d\}+0.5d=0.5d-0.5d=0,\\
\lim_{u\rightarrow+\infty}\mathrm{CL}_\sigma(u|d)&=&s(\sigma|d)\{\sigma\log(2)\}+0.5d\{1+s(\sigma|d)\}\\
&=&s(\sigma|d)\{\sigma\log(2)+0.5d\}+0.5d=0.5d+0.5d=d,
\end{eqnarray*}
and
\begin{eqnarray*}
\lim_{\sigma\rightarrow0^+}\mathrm{CL}_\sigma(u|d)&=&\max(u,0)-u+\{d-\max(d-u,0)\}\\
&=&\max(u,0)-u+\min(u,d)=\mathrm{CReLU}(u|d).
\end{eqnarray*}
Moreover, $\mathrm{CL}_\sigma(u|d)$ it is point symmetric in $(0.5d,0.5d)$. First, it is easy to verify that $\mathrm{CL}_\sigma(0.5d|d)=0.5d$. By (\ref{Clinear}), it is easy to verify that
\begin{eqnarray*}
\mathrm{CL}_\sigma(0.5d+v|d)-0.5d=-\{\mathrm{CL}_\sigma(0.5d-v|d)-0.5d\},\ \text{for}\ v\in[0,0.5d].
\end{eqnarray*}
Now, assuming $v\in(0.5d,+\infty)$, we have
\begin{eqnarray*}
&&\mathrm{CL}_\sigma(0.5d+v|d)-0.5d\\
&=&s(\sigma|d)\{\mathrm{L}_\sigma(0.5d+v)-(0.5d+v)-\mathrm{L}_\sigma(d-0.5d-v)+0.5d\}\\
&=&s(\sigma|d)\{\mathrm{L}_\sigma(0.5d+v)-v-\mathrm{L}_\sigma(0.5d-v)\}
\end{eqnarray*}
and
\begin{eqnarray*}
&&\mathrm{CL}_\sigma(0.5d-v|d)-0.5d\\
&=&s(\sigma|d)\{\mathrm{L}_\sigma(0.5d-v)-(0.5d-v)-\mathrm{L}_\sigma(d-0.5d+v)+0.5d\}\\
&=&s(\sigma|d)\{\mathrm{L}_\sigma(0.5d-v)+v-\mathrm{L}_\sigma(0.5d+v)\}.
\end{eqnarray*}
Thus,
\begin{eqnarray*}
\mathrm{CL}_\sigma(0.5d+v|d)-0.5d=-\{\mathrm{CL}_\sigma(0.5d-v|d)-0.5d\},\ \text{for}\ v\in(0.5d,+\infty)].
\end{eqnarray*}

The maximal deviation of $\mathrm{CL}_\sigma(u|d)$ to $\mathrm{CReLU}(u|d)$ is at $u = 0$ and $u = d$ (of opposite sign), and it
is of absolute size $0.5d\{1-s(\sigma|d)\}$.

The clipped Laplace link function $\mathrm{CL}_\sigma(u|d)$ not only serves as a mean function in regression modeling, but also relates to the uniform distribution. Note that $\mathrm{CReLU}(u|d)/d$ just equals the CDF of the uniform distribution $U(0,d)$. Since $\mathrm{CL}_\sigma(u|d)/d$ is itself a CDF and approaches $U(0,d)$ for $\sigma\rightarrow 0$, thus $\mathrm{CL}_\sigma(u|d)/d$ might be viewed as a `soft uniform distribution', whose support set is $\mathbb{R}$.

\section{The MVJ($p_1,p_2$) bounded $\mathbb{Z}$-valued model} \setcounter {equation}{0}
\def\theequation{\thesection.\arabic{equation}}

Define $\mu_t=E(D_t|\mathcal {F}_{t-1})$. We propose the following MVJ model
\begin{eqnarray}\label{Dt}
\left\{
  \begin{array}{l}
 D_t=X(\kappa_1(\mu_t,r_t,U_{1t}),\kappa_2(\mu_t,r_t,U_{2t}),U_{0t})\\
 \mu_t=\mathrm{CL}_\sigma(\xi_t |d),\\
\xi_t=\xi_t(\vtheta)=c+\sum_{i=1}^{p_1}\phi_iD_{t-i}+\sum_{j=1}^{p_2}\psi_j\mu_{t-j},
  \end{array}
\right.
\end{eqnarray}
where $p_1\geq1$, $p_2\geq0$, $\phi_{p_1},\psi_{p_2}\neq0$, $\vtheta=(c,\phi_1,\cdots,\phi_{p_1},\psi_1,\cdots,\psi_{p_2})^\textsf{T}$; $(U_{0t})$, $(U_{1t})$ and $(U_{2t})$ are three sequences of i.i.d. uniform random variables defined on $[0,1]$; $(r_{t})$ is a sequence of i.i.d. random variables with support set contained in $[0,1]$; $U_{0t}$, $U_{1t}$, $U_{2t}$, and $r_t$ are mutually independent; and $X(\cdot,\cdot,\cdot)$ is defined in (\ref{binary}). Let $\vvartheta=(\vartheta_1,\vartheta_2)^\textsf{T}$, where $\vartheta_k=E(r_t^k)$, for $k=1,2$. The unknown parameter vectors $\vtheta$ and $\vvartheta$ determine the conditional mean and variance structure of $D_t$. In model (\ref{Dt}), the distribution of $r_t$ is not specified and remains nonparametric. The clipped-Laplace link function satisfies that
\begin{eqnarray*}
\mathrm{CL}_\sigma(u|d)\leq 0.5d+L_0(u),\ u\in\mathbb{R},\ \ \sigma>0.
\end{eqnarray*}
where $L_0(u)=u^+=\max(u,0)$ is the ReLU function. Define
\begin{eqnarray*}
e_t&=&D_t-\mu_t\\
&=&\kappa_{1t}{1}\biggr(U_{0t}\leq \frac{\kappa_{2t}-\mu_t}{\kappa_{2t}-\kappa_{1t}}\biggr)+\kappa_{2t}{1}\biggr(U_{0t}> \frac{\kappa_{2t}-\mu_t}{\kappa_{2t}-\kappa_{1t}}\biggr)-\mu_t,
\end{eqnarray*}
where $\kappa_{1t}=\kappa_1(\mu_t,r_t,U_{1t})$ and $\kappa_{2t}=\kappa_2(\mu_t,r_t,U_{2t})$. Then, $|e_t|=|D_t-\mu_t|\leq |\kappa_{1t}-\mu_t|+|\kappa_{2t}-\mu_t|\leq \mu_t+d-\mu_t=d$.

\begin{remark}\label{meanspace}
Let $\vtheta=(c,\phi_1,\cdots,\phi_{p_1},\psi_1,\cdots,\psi_{p_2})^\textsf{T}$. For $\mu_t=\mathrm{CL}_\sigma(\xi_t|d)$ $(\sigma\geq0)$, we consider two parameter spaces: $\Theta_0=\{\vtheta: c\in\mathbb{R}, \sum_{i=1}^{p_1}|\phi_i|+\sum_{j=1}^{p_2}|\psi_j|<1\}$ and $\Theta_1=\{\vtheta: c, \phi_1, ..., \phi_{p_1}, \psi_1, ..., \psi_{p_2}\geq0, c+\sum_{i=1}^{p_1}\phi_i+\sum_{j=1}^{p_2}\psi_j<1\}$. Obvioulsy, $\Theta_1\subset \Theta_0$.
If $\vtheta\in\Theta_1$, then $\mu_t=\mathrm{CL}_\sigma(c+\sum_{i=1}^{p_1}\phi_iD_{t-i}+\sum_{j=1}^{p_2}\psi_j\mu_{t-j}|d)\equiv s(\sigma|d)\{c+\sum_{i=1}^{p_1}\phi_iD_{t-i}+\sum_{j=1}^{p_2}\psi_j\mu_{t-j}\}+0.5d\{1-s(\sigma|d)\}$, which is a linear function of $(D_{t-1},D_{t-2},\cdots)$. For $\mu_t=\mathrm{CS}_\sigma(\xi_t|d)$ $(\sigma\geq0)$, we consider the parameter space $\Theta_0$. For all $\vtheta\in\Theta_0$, $\mu_t=\mathrm{CS}_\sigma(\xi_t|d)$ is always a nonlinear function of $(D_{t-1},D_{t-2},\cdots)$.
\end{remark}
Let $\mathscr{F} = (\mathcal {F}_T)$, $T\geq0$ be the natural filtration associated to the MVJ($p_1,p_2$) process, where $\mathcal {F}_T=\sigma((U_{0t},U_{1t},U_{2t},r_t),t\leq T)$. Then, the study of the MVJ($p_1,p_2$) process can be carried out through the following vectorized process
\begin{eqnarray}\label{vinarp}
\Y_t=\left(
   \begin{array}{c}
D_t\\
\mu_{t}\\
\vdots\\
D_{t-p+1}\\
\mu_{t-p+1}
\end{array}\right)=\left(
   \begin{array}{c}
X(\kappa_1(\mu_t,r_t,U_{1t}),\kappa_2(\mu_t,r_t,U_{2t}),U_{0t})\\
\mu_{t}\\
\vdots\\
D_{t-p+1}\\
\mu_{t-p+1}
\end{array}\right),
\end{eqnarray}
where $p=\max(p_1,p_2)$. The process $(\Y_t)$ forms a homogeneous Markov chain with state space $\mathbb{E} =(\{0,\cdots,d\}\times[0,d])^p$. For $\x=(x_1,\cdots, x_{2p})^\textsf{T}\in\mathbb{E}$ and $\y=(y_1,\cdots, y_{2p})^\textsf{T}\in\mathbb{E}$,  the transition probability
function from $\x$ to $\y$ is given by
\begin{eqnarray*}
\pi(\x,\y)=\mathbb{P}\{y_1=X(\kappa_1(\mu_1,r_1,U_{11}),\kappa_2(\mu_1,r_1,U_{21}),U_{01})\}{1}(y_2=\mu_1,y_3=x_1,...,y_{2p}=x_{2p-2}),
\end{eqnarray*}
where $\mu_1=\mathrm{CL}_\sigma(c+\sum_{i=1}^{p}\phi_ix_{2i-1}+\sum_{j=1}^{p}\psi_jx_{2j}|d)$, $\phi_{i}=0$ for $p_1<i\leq p$, and $\psi_{j}=0$ for $p_2<j\leq p$.

\begin{proposition}\label{ergodicity}
Let $\psi(z)=1-\sum_{j=1}^{p_2}\psi_jz^i$ and $\varphi_*(z)=1-\sum_{j=1}^p(\phi_{j}^+ +\psi_{j}^+)z^{j}$, where $p=\max(p_1,p_2)$, $a^+=\max\{a,0\}$, $\phi_{j}^+=0$ for $p_1<j\leq p$, and $\psi_{j}^+=0$ for $p_2<j\leq p$. Suppose that:
\begin{itemize}
\item[1.]The Markov chain $(\Y_t)$ is irreducible and aperiodic;
\item[2.]$\psi(z)\neq0$, $\varphi_*(z)\neq0$, for $z\in\mathbb{C}$ and $|z|\leq1$;
\end{itemize}
Then
\begin{itemize}
\item[1.]The MVJ$(p_1,p_2)$ process $(\Y_t)$ has a unique invariant probability measure $\lambda$ which has a moment of any order $k$ (i.e., $\lambda(\|\cdot\|^k) <+\infty$).;
\item[2.]For all $\y \in \mathbb{E}$ and $g \in L_1(\lambda)$, we have
$
\frac{1}{T}\sum_{t=1}^Tg(\Y_t)\longrightarrow \lambda(g),\ \
\mathbb{P}_{\text{\y}}\ a.s.
$
where $\mathbb{P}_{\text{\y}}$ denotes the conditional probability
$\mathbb{P}(\cdot)=\mathbb{P}(\cdot|\Y_0=\y)$.
\end{itemize}
\end{proposition}

\subsection{Autocorrelation structure}

Let $c_\star=c s(\sigma|d)+0.5d\{1-s(\sigma|d)\}$, $\phi_{\star i}=s(\sigma|d)\phi_i$, and $\psi_{\star j}=s(\sigma|d)\psi_j$, where $i=1,\cdots,p_1$ and $j=1,\cdots,p_2$. If $\vtheta\in\Theta_1$, by (\ref{Clinear}) and the definition of $e_t$, we can write
\begin{eqnarray*}
D_t&=&\mu_t+e_t\\
&=&cs(\sigma|d)+0.5d\{1-s(\sigma|d)\}+\sum_{i=1}^{p_1}s(\sigma|d)\phi_iD_{t-i}+\sum_{j=1}^{p_2}s(\sigma|d)\psi_j\mu_{t-j}+e_t\\
&=&c_\star+\sum_{i=1}^{p_1}\phi_{\star i}D_{t-i}+\sum_{j=1}^{p_2}\psi_{\star j}\mu_{t-j}+e_t\\
&=&c_\star+\sum_{i=1}^{p_1}\phi_{\star i}D_{t-i}+\sum_{j=1}^{p_2}\psi_{\star j}D_{t-j}+e_t-\sum_{j=1}^{p_2}\psi_{\star j}e_{t-j}\\
&=&c_\star+\{\phi_{\star}(B)+\psi_{\star}(B)\}D_{t}+e_t+\delta_{\star}(B)e_{t}\\
&=&c_\star+\varphi_{\star}(B)D_{t}+\{1+\delta_{\star}(B)\}e_{t},
\end{eqnarray*}
where $\phi_{\star}(B)=\sum_{i=1}^{p_1}\phi_{\star i}B^i$, $\psi_{\star}(z)=\sum_{j=1}^{p_2}\psi_{\star j}B^j$, $\varphi_{\star}(B)=\sum_{k=1}^{p}\varphi_{\star k}B^k=\sum_{i=1}^{p_1}\phi_{\star i}B^i+\sum_{j=1}^{p_2}\psi_{\star j}B^j$, $p=\max\{p_1,p_2\}$, and $\delta_{\star}(B)=\sum_{k=1}^{p_2}\delta_{\star k}B^k=\sum_{j=1}^{p_2}(-\psi_{\star j})B^j$ with $\delta_{\star k}=-\psi_{\star k}$. It follows that
\begin{eqnarray*}
D_t&=&\frac{c_\star}{1-\varphi_{\star}(1)}+\{1-\varphi_{\star}(B)\}^{-1}\{1+\delta_{\star}(B)\}e_{t}=\frac{c_\star}{1-\varphi_{\star}(1)}+\sum_{i=0}^{+\infty}\varpi_ie_{t-i},
\end{eqnarray*}
where $\varpi(B)=\{1-\varphi_{\star}(B)\}^{-1}\{1+\delta_{\star}(B)\}=\sum_{i=0}^{+\infty}\varpi_iB_i$ with $\varpi_0=1$.

Let $\mathscr{F} = (\mathcal{F}_T)_{T\in\mathbb{Z}}$ be the natural
filtration associated to the MVJ$(p_1,p_2)$ process where
$\mathcal{F}_T =\sigma((U_{0t},U_{1t},U_{2t},r_t), t \leq T)$
for $n\in\mathbb{Z}$, and $\mathcal{F}_{-\infty}$ is the degenerated
$\sigma$-algebra. By (\ref{cmuxr}) and (\ref{cvarxr}), we get
$E(D_t|\mathcal{F}_{t-1})=\mu_t$ and
$\var(D_t|\mathcal{F}_{t-1})=R(\mu_t)+V(\mu_t)$, where $V(\mu)=\vartheta_2\Delta(\mu)\{d-\Delta(\mu)-1\}
+\vartheta_1\bigr[\{\mu-\Delta(\mu)\}\{d-\Delta(\mu)-1\}+\Delta(\mu)\{\Delta(\mu)+1-\mu\}\bigr]\geq0$.
The following proposition gives the autocorrelation structure of the
MVJ$(p_1,p_2)$ process when $\vtheta\in\Theta_1$.

\begin{proposition}\label{autocorrelation}
Suppose that $\vtheta\in\Theta_1$ and the conditions of Proposition \ref{ergodicity} are
satisfied. Let
\begin{eqnarray*}
&&E(D_t)=\mu,\ \ \mbox{for all}\ t,\\
&&E(D_t-\mu)(X_{t-j}-\mu)=\gamma_j,\ \ \mbox{for all}\ t\ \mbox{and}\ j,\\
&&\rho_j=\gamma_j/\gamma_0,\ \ j\in\mathbb{Z}_+,
\end{eqnarray*}
then we have
\begin{eqnarray*}
\mu=\frac{c_\star}{1-\sum_{i=1}^{p_1}\phi_{\star i}-\sum_{j=1}^{p_2}\psi_{\star j}}\ \ \text{and}\ \ \gamma_k=E(e_{t}^2)\sum_{i=0}^{+\infty}\varpi_i\varpi_{k+i},\ \ k\geq0,
\end{eqnarray*}
where $E(e_{t}^2)=E\{R(\mu_t)+V(\mu_t)\}$ and $(\varpi_i)_{i\geq0}$ satisfies $\varpi(B)=\{1-\varphi_{\star}(B)\}^{-1}\{1+\delta_{\star}(B)\}=\sum_{i=0}^{+\infty}\varpi_iB_i$ with $\varpi_0=1$.
\end{proposition}

\subsection{Prediction}

For the MVJ($p_1,p_2$) model, using (\ref{cmuxr}) and (\ref{cvarxr}), the one-step predictors of mean and
variance are given respectively by
\begin{eqnarray*}
E(D_{T+1}|\mathcal {F}_{T})=\mu_{T+1}\ \ \text{and}\ \ \var(D_{T+1}|\mathcal{F}_{T})=R(\mu_{T+1})+h_{T+1},
\end{eqnarray*}
where
\begin{eqnarray*}
&&h_{T+1}=V(\mu_{T+1})=\vartheta_2\Delta(\mu_{T+1})\{d-\Delta(\mu_{T+1})-1\}\nonumber\\
&&+\vartheta_1\bigr[\{\mu_{T+1}-\Delta(\mu_{T+1})\}\{d-\Delta(\mu_{T+1})-1\}+\Delta(\mu_{T+1})\{\Delta(\mu_{T+1})+1-\mu_{T+1}\}\bigr],
\end{eqnarray*}
$\mu_{T+1}=\mathrm{CL}_\sigma(\xi_{T+1}(\vtheta) |d)$, $\xi_t(\vtheta)$ is given in (\ref{Dt}), and $\vartheta_k=E(r_t^k)$, for $k=1,2$.

The one-step predictors of mean and variance depend on the unknown parameter vectors $\vtheta$ and $\vvartheta$.  We will give consistent estimators of these parameter vectors in the next subsection.

\subsection{Conditional least-squares estimation}

For any generic $\vtheta=(c,\phi_1,\cdots,\phi_{p_1},\psi_1,\cdots,\psi_{p_2})^\textsf{T}\in\Theta_0$, define
\begin{eqnarray}\label{muyx}
\mu_{t}=\mu_{t}(\vtheta)=\mathrm{CL}_\sigma\biggr(c+\sum_{i=1}^{p_1}\phi_iD_{t-i}+\sum_{j=1}^{p_2}\psi_j\mu_{t-j}(\vtheta)|d\biggr),
\end{eqnarray}
and
\begin{eqnarray}\label{varyx}
&&h_{t}=h_{t}(\vvartheta|\vtheta)\nonumber\\
&=&\vartheta_1V_{1t}(\mu_t)+\vartheta_2V_{2t}(\mu_t)\nonumber\\
&=&\vartheta_2\Delta(\mu_t)\{d-\Delta(\mu_t)-1\}\nonumber\\
&&+\vartheta_1\bigr[\{\mu_t-\Delta(\mu_t)\}\{d-\Delta(\mu_t)-1\}+\Delta(\mu_t)\{\Delta(\mu_t)+1-\mu_t\}\bigr],
\end{eqnarray}
where $\vvartheta=(\vartheta_1,\vartheta_2)^\textsf{T}$, $\vartheta_1=E(r_t)$, $\vartheta_2=E(r_t^2)$, $V_{1t}=V_1(\mu_t)=\{\mu_t-\Delta(\mu_t)\}\{d-\Delta(\mu_t)-1\}+\Delta(\mu_t)\{\Delta(\mu_t)+1-\mu_t\}\geq0$, and $V_{2t}=V_2(\mu_t)=\Delta(\mu_t)\{d-\Delta(\mu_t)-1\}\geq0$. Let $(W_t)$ be a stationary sequence of positive weights such that
$W_t\in\sigma(D_{t-1},\mu_{t-1},\cdots)$. The WLS estimator of $\vtheta$ is defined as
\begin{eqnarray}\label{cls1}
\hat{\vtheta}_{T}=\arg\min_{\text{\vtheta}\in\Theta}\sum_{t={1}}^TW_t\{D_t- \mu_t(\vtheta)\}^2,
\end{eqnarray}
where $\vtheta\in\Theta$ and $\Theta$ is a compact set of $\Theta_0$. If $W_t\equiv1$, $\hat{\vtheta}_T$ is just the ordinary least squares (OLS)  estimator of $\vtheta$. In practice, minimization of (\ref{cls1}) can be done by an approximation procedure. Let $D_{t}=\tilde{\mu}_{t}(\vtheta)=0$ for $t\leq0$. Then, $\mu_t(\vtheta)$ can be approximated by
\begin{eqnarray*}\label{wls1approximate0}
\tilde{\mu}_t(\vtheta)=M\biggr(c+\sum_{i=1}^{p_1}\phi_iX_{t-i}+\sum_{j=1}^{p_2}\psi_j\tilde{\mu}_{t-j}(\vtheta)\biggr),\ t\in\mathbb{Z}.
\end{eqnarray*}
Correspondingly, $\hat{\vtheta}_T$ can be approximated by the solution of
$
\arg\min_{\text{\b}\in\Theta}\frac{1}{T}\sum_{t=1}^TW_t\left\{D_t-\tilde{\mu}_{t}(\b)\right\}^2.
$
We have the following results:
\begin{theorem}\label{consistency}
Suppose that the conditions of Proposition \ref{ergodicity} are
satisfied. Furthermore, assume (i) $\mathbb{P}\{\mu_t(\vtheta)=\mu_t(\b)\}=1$ implies that $\vtheta=\b$; (ii) $E(W_t)<+\infty$.
Then, $\hat{\vtheta}_T$ is strongly consistent, i.e.
$\hat{\vtheta}_T\asrow\vtheta$.
\end{theorem}

\begin{theorem}\label{norm}
Suppose that the conditions of Theorem \ref{consistency} are
satisfied. Assume furthermore that the sequence of weights $(W_t)$
satisfies
\begin{eqnarray*}
&&E\biggr[\sup_{\text{\b}\in\mathcal {B}}\biggr\|W_t^{1/2}\frac{\partial \mu_{t}(\b)}{\partial \b}\biggr\|^2\biggr]<\infty,\ \ \lambda_{\min}\biggr[E\biggr\{W_t\frac{\partial \mu_{t}(\vtheta)}{\partial \vtheta}\frac{\partial \mu_{t}(\vtheta)}{\partial \vtheta^\textsf{T}}\biggr\}\biggr]>0,\\
&&E\biggr[\sup_{\text{\b}\in\Theta}\biggr\|W_t\{D_t-\mu_{t}(\b)\}\frac{\partial \mu_{t}(\b)}{\partial \b}\biggr\|^2\biggr]<\infty,\ \text{and}\ E\biggr\{W_t\biggr\|\frac{\partial \mu_{t}(\text{\vtheta})}{\partial \text{\vtheta}}\biggr\|^2\biggr\}<+\infty,
\end{eqnarray*}
where $\mathcal {B}\subseteq \Theta$ is a neighborhood of $\vtheta$ and $\lambda_{\min}(A)$ is the smallest eigenvalue of the matrix $A$. Then, the WLS estimator $\hat{\vtheta}_T$ is asymptotically normal,
i.e.
\begin{eqnarray*}
\sqrt{T}(\hat{\vtheta}_T-\vtheta)\longrightarrow
N(0,K_1^{-1}\Gamma_1 K_1^{-1}),
\end{eqnarray*}
where
\begin{eqnarray*}
K_1=E\biggr\{W_t\frac{\partial \mu_{t}(\vtheta)}{\partial \vtheta}\frac{\partial \mu_{t}(\vtheta)}{\partial \vtheta^\textsf{T}}\biggr\},\ \ \ \ \
\Gamma_1=E\biggr[W_t^2\{D_t-\mu_{t}(\vtheta)\}^2\frac{\partial \mu_{t}(\vtheta)}{\partial \vtheta}\frac{\partial \mu_{t}(\vtheta)}{\partial \vtheta^\textsf{T}}\biggr].
\end{eqnarray*}
\end{theorem}

\begin{remark}
To ensure that the matrices $K_1$ and $\Gamma_1$ are well defined in WLS estimation, the integrability conditions on the weights $(W_t)$ are crucial. The optimal choice of these weights in WLS estimation is typically given by:
\begin{eqnarray}\label{conditionvar}
W_t=Var(D_t|\mathcal
{F}_{t-1})^{-1}=\{R(\mu_t)+h_t\}^{-1},\
\ t\in\mathbb{Z}_+
\end{eqnarray}
where $\mu_t=\mu_t(\vtheta)$ and $h_{t}=h_{t}(\vvartheta|\vtheta)$ are given by (\ref{muyx}) and (\ref{varyx}).
\end{remark}
To obtain feasible optimal weights $(W_t)$ in (\ref{conditionvar}), which depend on the unknown parameter vector $(\vtheta^\textsf{T},\vvartheta^\textsf{T})$, it is necessary to follow a two-step procedure. First, an estimate (e.g., the OLS estimate) of $(\vtheta^\textsf{T},\vvartheta^\textsf{T})$ needs to be plugged into $(W_t)$ in (\ref{conditionvar}). Then, one uses feasible optimal weights to construct the optimal WLS (OWLS) estimator of $\vtheta$. The following theorem justifies this process:
\begin{theorem}\label{twostep}
Let
$(\hat{\vtheta}^\textsf{T},\hat{\vvartheta}^\textsf{T})$ be a sequence of estimators such that
$\sqrt{T}((\hat{\vtheta}-\vtheta)^\textsf{T},(\hat{\vvartheta}-\vvartheta)^\textsf{T})=O_{\mathbb{P}}(1)$.
Suppose that the conditions of Proposition \ref{ergodicity} are satisfied. Define the OWLS estimator of $\vtheta$ as
\begin{eqnarray}\label{wlsoptimal}
\breve{\vtheta}_T=\arg\min_{\text{\b}\in\Theta}\frac{1}{T}\sum_{t=1}^T\hat{W}_t\left\{D_t-\mu_{t}(\b)\right\}^2.
\end{eqnarray}
where
$\hat{W}_t=\left(R(\hat{\mu}_t)+\hat{h}_t\right)^{-1}$ with
$\hat{h}_{t}=h_{t}(\hat{\vvartheta}|\hat{\vtheta})$ and $\hat{\mu}_t=\mu_t(\hat{\vtheta})$.
Then, we have
$
\sqrt{T}(\breve{\vtheta}_T-\vtheta)\longrightarrow N(0,\Sigma),
$
where
$$
\Sigma^{-1}=E\left\{\left(R(\mu_t)+h_t\right)^{-1}\frac{\partial \mu_{t}(\vtheta)}{\partial \vtheta}\frac{\partial \mu_{t}(\vtheta)}{\partial \vtheta^\textsf{T}}\right\}.
$$
\end{theorem}
Define $\hat{\varepsilon}_t=D_t-\hat{\mu}_t$, where $\hat{\mu}_t=\mathrm{CL}_\sigma(\xi_t(\hat{\vtheta}_{OLS})|d)$ and $\hat{\vtheta}_{OLS}=\arg\min_{\text{\vtheta}\in\Theta}\sum_{t=1}^T\{D_t- \mu_t(\vtheta)\}^2$. Then, the OLS estimator of $\vvartheta=(\vartheta_1,\vartheta_2)^\textsf{T}$ is defined as
\begin{eqnarray}\label{cls2}
\hat{\vvartheta}_{T}=\arg\min_{\text{\vvartheta}\in\Lambda}\sum_{t={1}}^T\{\hat{\varepsilon}_t^2-R(\hat{\mu}_t)-\hat{\V}_t^\textsf{T}\vvartheta\}^2,
\end{eqnarray}
where $\Lambda$ is a compact set of $[0,1]\times[0,1]$, $\hat{\V}_t=(\hat{V}_{1t},\hat{V}_{2t})^\textsf{T}$ and $\hat{V}_{kt}=V_k(\hat{\mu}_t)$, $k=1,2$. It is easy to verify that
\begin{eqnarray}\label{cls3}
\hat{\vvartheta}_{T}=\biggr(\sum_{t={1}}^T\hat{\V}_t\hat{\V}_t^T\biggr)^{-1}\sum_{t={1}}^T\hat{\V}_t\{\hat{\varepsilon}_t^2-R(\hat{\mu}_t)\}.
\end{eqnarray}
We have the following result:
\begin{theorem}\label {vartheconsist}
Suppose that the conditions of Theorem \ref{norm} are satisfied. Furthermore, assume that $\mathbb{P}\{h_t(\vvartheta|\vtheta)=h_t(\s|\vtheta)\}=1$ implies that $\vvartheta=\s$. Then,
$\hat{\vvartheta}_T$ is weakly consistent, i.e. $\hat{\vvartheta}_T\prow \vvartheta$.
\end{theorem}

\begin{theorem}\label {varthenormal}
Suppose that the conditions of Theorem \ref{vartheconsist} are
satisfied. Let $u_t(\vvartheta|\vtheta)=\{D_t-\mu_t(\vtheta)\}^2-R(\mu_t(\vtheta))-h_t(\vvartheta|\vtheta)$. Furthermore, assume that
\begin{eqnarray*}
&&E\biggr\{\sup_{\text{\b}\in\mathcal {B},\s\in\Lambda}\biggr\|\frac{\partial h_t(\text{\s}|\text{\b})}{\partial\text{\s}}\biggr\|^2\biggr\}<+\infty,\ \  E\biggr\{\sup_{\text{\b}\in\mathcal {B},\s\in\Lambda}\biggr\|\frac{\partial h_t(\text{\s}|\text{\b})}{\partial\text{\s}}\biggr\|\biggr\|\frac{\partial u_t(\text{\s}|\text{\b})}{\partial \text{\b}^\textsf{T}}\biggr\|\biggr\}<+\infty,\\
&&\lambda_{\min}\biggr[E\biggr\{\frac{\partial h_t(\vvartheta|\vtheta)}{\partial\vvartheta}\frac{\partial u_t(\vvartheta|\vtheta)}{\partial \vvartheta^\textsf{T}}\biggr\}\biggr]>0,\ \text{and}\ E\biggr\{\biggr\|\frac{\partial h_t(\vvartheta|\vtheta)}{\partial\vvartheta}u_t(\vvartheta|\vtheta)\biggr\|^2\biggr\}<+\infty.
\end{eqnarray*}
Then, the OLS
estimator $\hat{\vvartheta}_T$ is asymptotically normal, i.e.
\begin{eqnarray*}
\sqrt{T}(\hat{\vvartheta}_T-\vvartheta)\longrightarrow
N(0,\Omega),
\end{eqnarray*}
where $\Omega$ is given in the proof of this theorem.
\end{theorem}
The algorithms for the computations of the OLS and OWLS estimates and their estimated asymptotic covariance matrices
are provided in the Section S2 of the Supplementary Material.

\subsection{Model selection and model diagnostics}
In this subsection, we consider the model selection problem for the
MVJ models. We face the challenge that the true conditional likelihood for the MVJ model can't be directly computed due to the unspecified distribution of $r_t$ in (\ref{Dt}). To navigate this issue, we consider an approach similar to what was proposed by Hurvich \& Tsai (1995) and later adapted by Liu \& Yuan (2013), which involves employing a conditional Gaussian quasi-likelihood method. This method allows for the construction of both the Akaike Information Criterion (AIC) and the Bayesian Information Criterion (BIC) for model selection, even when the true likelihood is not available.

  The resultant AIC and BIC are given respectively as
\begin{eqnarray*}
\mbox{AIC}(p_1,p_2)&=&T\log\left\{T^{-1}\sum_{t=1}^T(y_t-\hat{\mu}_t)^2 \right\}+2(3+p_1+p_2),
\end{eqnarray*}
and
\begin{eqnarray*}
\mbox{BIC}(p_1,p_2)&=&T\log\left\{T^{-1}\sum_{t=1}^T (y_t-\hat{\mu}_t)^2\right\}+\log(T-p-1)(3+p_1+p_2),
\end{eqnarray*}
where $p=\max\{p_1,p_2\}$, $\hat{\mu}_t=\mathrm{CL}_\sigma(\xi_t(\hat{\vtheta}_{OLS}) |d)$, $\hat{\vtheta}_{OLS}$ is the OLS estimator of $\vtheta$, and $\xi_t(\vtheta)$ is defined by (\ref{Dt}).

Let $(p_{1m},p_{2m})$ be a maximum model order cut-offs for the MVJ$(p_1,p_2)$
model. The selected order using AIC and BIC is given respectively by
\begin{eqnarray*}
(\hat{p}_1,\hat{p}_2)_{AIC}=\arg\min_{p_k\leq p_{km},k=1,2}\mbox{AIC}(p_1,p_2),
\end{eqnarray*}
and
\begin{eqnarray*}
(\hat{p}_1,\hat{p}_2)_{BIC}=\arg\min_{p_k\leq p_{km},k=1,2}\mbox{BIC}(p_1,p_2).
\end{eqnarray*}
We investigate the performances of these two criteria for selecting the MVJ
models through simulations in the next section.

\subsection{Model diagnostics}

To check the adequacy of conditional mean and variance assumptions, it is common to use the standardized Pearson residuals $(r_t)$, which are defined as
\begin{eqnarray*}
r_t(\vtheta_T,\vvartheta_T)=\frac{X_t-E(X_t|\mathcal {F}_{t-1};\vtheta_T,\vvartheta_T)}{\sqrt{\var(X_t|\mathcal {F}_{t-1};\vtheta_T,\vvartheta_T)}}=\frac{X_t-\mu_t(\vtheta_T)}{\sqrt{R(\mu_t(\vtheta_T))+\vartheta_{1T}V_{1}(\mu_t(\vtheta_T))+\vartheta_{2T}V_{2}(\mu_t(\vtheta_T))  }},
\end{eqnarray*}
where $\vvartheta_T=(\vartheta_{1T},\vartheta_{2T})^\textsf{T}$ is the OLS estimator of $\vvartheta$ and $\vtheta_T$ is the OLS or OWLS estimator of $\vtheta$. For an adequate model, the Pearson residuals should meet the specified criteria: mean zero, variance one, and uncorrelated.

To assess the adequacy of the dispersion structure, Aknouche and Scotto (2024) utilized two metrics: the mean absolute residual (MAR), $\text{MAR}=\frac{1}{T}\sum_{t=1}^T|X_t-\mu_t(\vtheta_T)|$, and the mean squared Pearson residual, $\text{MSPR}=\frac{1}{T}\sum_{t=1}^Tr_t^2(\vtheta_T,\vvartheta_T)$. A smaller value of MAR and $|\text{MSPR}- 1|$ indicates a better fit of the model to the data, suggesting that the model adequately captures the dispersion structure.

\section{Simulation} \setcounter {equation}{0}
\def\theequation{\thesection.\arabic{equation}}
To evaluate the efficiency of the OLS and OWLS estimators and the performances of AIC and BIC for
selecting the MVJ models, we conduct two simulation studies.
All simulations are carried out in the R Project for Statistical Computing. Moreover, in all
simulations, we set $d=15$, and the innovation variable, say $r_t$, is generated from a Beta
distribution with parameter $(1,1)$. Additional simulation results with different Beta distributions of $r_t$ presented in the Section S3 of the Supplementary Material yield similar
results.

For the Monte Carlo studies, we consider the following two settings:
\begin{itemize}
\item[]Setting (a):
\item[$M_1$:] MVJ(1,0) model with   $(\vtheta^\textsf{T},\vvartheta^\textsf{T} )=(-0.2,0.5,1/2,1/3)$;
\item[$M_2$:] MVJ(1,1) model with  $(\vtheta^\textsf{T},\vvartheta^\textsf{T} )=(-0.2,0.4,0.4,1/2,1/3)$;
\item[$M_3$:] MVJ(1,2) model with  $(\vtheta^\textsf{T},\vvartheta^\textsf{T} )=(-0.2,0.4,0.1,0.4,1/2,1/3)$;
\item[$M_4$:] MVJ(2,0) model with    $(\vtheta^\textsf{T},\vvartheta^\textsf{T} )=(-0.2,0.2,0.5,1/2,1/3)$;
\item[$M_5$:] MVJ(2,1) model with   $(\vtheta^\textsf{T},\vvartheta^\textsf{T} )=(-0.2,0.1,0.4,0.4,1/2,1/3)$;
\item[$M_6$:] MVJ(2,2) model with   $(\vtheta^\textsf{T},\vvartheta^\textsf{T} )=(-0.2,0.1,0.4,0.1,0.3,1/2,1/3)$;
\end{itemize}
\begin{itemize}
\item[]and Setting (b):
\item[$M_1$:] MVJ(1,0) model with   $(\vtheta^\textsf{T},\vvartheta^\textsf{T} )=(5,-0.5,1/2,1/3)$;
\item[$M_2$:] MVJ(1,1) model with  $(\vtheta^\textsf{T},\vvartheta^\textsf{T} )=(5,-0.4,-0.4,1/2,1/3)$;
\item[$M_3$:] MVJ(1,2) model with  $(\vtheta^\textsf{T},\vvartheta^\textsf{T} )=(5,-0.4,-0.1,-0.4,1/2,1/3)$;
\item[$M_4$:] MVJ(2,0) model with    $(\vtheta^\textsf{T},\vvartheta^\textsf{T} )=(5,-0.2,-0.5,1/2,1/3)$;
\item[$M_5$:] MVJ(2,1) model with   $(\vtheta^\textsf{T},\vvartheta^\textsf{T} )=(5,-0.1,-0.4,-0.4,1/2,1/3)$;
\item[$M_6$:] MVJ(2,2) model with   $(\vtheta^\textsf{T},\vvartheta^\textsf{T} )=(5,-0.1,-0.4,-0.1,-0.3,1/2,1/3)$.
\end{itemize}

Obviously, the setting (a) includes normal parameter scenarios while the setting (b) contains extreme parameter scenarios. To gain insight into the capabilities of MVJ models employing the clipped-Laplace link function $\mathrm{CL}_1(\cdot|d)$  in explaining various autocorrelation structures, we present sample ACF pairs
$(\rho_{X_t}(1),\rho_{X_t}(2))$ for the two different settings in Table 1. From Table 1, it's evident that both negative and non-negative ACFs can be accommodated by the MVJ models using the clipped-Laplace link function.

In the first simulation study, we employ the root mean squared error (RMSE) to assess the finite sample behavior of the conditional least squares estimators. We conduct $1000$ replications for each scenario. Two sample sizes are considered: $T=200$ and $T=500$. Simulation results for the settings (a) and (b) are
presented in Tables 2-3, respectively. The reported results show that the OWLS estimator $\breve{\vtheta}_T$ generally exhibits smaller RMSEs for $\vtheta$ compared to the OLS estimator $\hat{\vtheta}_T$ in most cases under setting (a). However, under setting (b), the OWLS estimator $\breve{\vtheta}_T$ does not demonstrate obvious efficiency gains over the OLS estimator $\hat{\vtheta}_T$.

In the second Monte Carlo study, we explore the performances of AIC and BIC for selecting the MVJ models utilizing the clipped-Laplace link function $\mathrm{CL}_1(\cdot)$. We consider a set of candidate models $\{M_j: j=1,\cdots,6\}$ under the setting (a) or (b). We assess the efficiency of AIC and BIC in selecting the most appropriate model from the candidate set.

We consider three sample sizes: $T=200$, $500$, and $800$. Tables 4-5 present the numbers of orders selected by AIC and BIC across 1000 realizations. From the results in Tables 4-5, we observe the following: For $T=200$ and $500$, AIC generally outperforms BIC, favoring larger models. However, for $T=800$, BIC tends to perform better, favoring simpler models. As the sample size increases from $T=200$ to $800$, both criteria show a higher probability of selecting the true model. This indicates a clear impact of sample size on the selection behavior of AIC and BIC, where AIC tends to prefer larger models and BIC tends to favor simpler ones.

\section{Applications to real data}
\setcounter {equation}{0}
\def\theequation{\thesection.\arabic{equation}}
To demonstrate how the MVJ models work, they are applied to
two data sets in completely different areas.

\subsection{Geyser eruption data}

In this subsection, we consider the Geyser eruption data analyzed by Jentsch \& Reichmann (2019) and Wei${\ss}$ \& Jahn (2022). These time series data refer to successive eruptions of the Old
Faithful Geyser and are provided through the command \verb"geyser" of the R-package \textsc{MASS}.  The length of this time series is $299$ and some of the observations do not belong to $\mathbb{Z}_+$.
To apply the bounded count time series model, Jentsch \& Reichmann (2019) and Wei${\ss}$ \& Jahn (2022) set $D_t=1$ ($D_t=0$) if the $t$-th eruption duration $S_t$ was at least (less than) three minutes. Here, we adopt a new discretization method $D_t=\Delta(S_t)$, which can retain more information about $\{S_t\}_{t=1}^{299}$.

Figure 4 (a) presents the plot of the resulted Geyser eruption data $\{D_t\}_{t=1}^{299}$, which exhibit strong negative autocorrelations with significant sample PACF
values for lags $\leq2$, see Figure 4 (b, c). The sample ACF and PACF imply that the MVJ(2,0) model should be considered. Moreover, $D_t\in\{0,1,2,3,4,5\}$. Thus, $d=5$.

We use the first $T=249$ observations for fitting the MVJ models with the clipped-Laplace link function $\mathrm{CL}_1(\cdot)$ and leave out the last $T_{new}=50$ observations for
a later forecast experiment. For model selection, we consider MVJ$(p_1,p_2)$ models with $(p_1,p_2)=(1,0)$, $(1,1)$, $(1,2)$, $(2,0)$, $(2,1)$, and $(2,2)$ as candidate models. The AIC and BIC values of the candidate models for the Geyser eruption counts are given in Table 6. It is easy to see that both of these two criteria select the
MVJ(2,0) model.

The final estimates together with their estimated standard deviations (SDs) are summarized in Table 7.  Obviously, the OWLS estimator $\breve{\vtheta}_T\notin \Theta_1$
and the OLS estimator $\hat{\vtheta}_T\notin \Theta_1$, which implies that the fitted MVJ(2,0) model has a nonlinear mean structure. Moreover, the OWLS estimator $\breve{\vtheta}_T$ gives smaller SDs for $\vtheta$ than the OLS estimator $\hat{\vtheta}_T$. Figure 4(d) presents the standardized residuals plot using the OLS estimates of $\vtheta$ and $\vvartheta$.

To check the adequacy of the fitted models, we consider the approaches discussed in Section 3.5. The model diagnostics statistics MAR and MSPR with the OLS and OWLS estimates of $\vtheta$ are presented in Table 8. The sample means ($E_T(r_t)$), sample SDs ($Sd_T(r_t)$), and the maximum absolute value of the sample autocorrelation ($\max_{1\leq k\leq 20}|\rho_{r_t}(k)|$) of the standardized residuals using the OLS and OWLS estimates of $\vtheta$ are also reported. From Table 8, we see that, the OLS and OWLS methods have the same MAR values but the MSPR value of the OLS method is closer to 1 than that of the OWLS method using the MVJ(2,0) model, which indicates a better model fit.

Finally, let us analyze the forecast performance of the fitted MVJ(2,0) model with the OLS and OWLS estimation methods. We apply the MAR and MSPR criteria to the 50 new Geyser eruption counts. The sample means ($E_{T_{new}}(r_t)$), sample SDs ($Sd_{T_{new}}(r_t)$), and the maximum absolute value of the sample autocorrelation ($\max_{1\leq k\leq 16}|\rho_{r_t}(k)|$) of the standardized prediction residuals using the OLS and OWLS estimates of $\vtheta$ are also considered. Results are summarized in Table 9.  Obviously, the OWLS-fitted MVJ(2,0) model shows the better predictive performance regarding the 50 Geyser eruption counts, according to the MAR criterion.

\subsection{Air quality data}

In this subsection, we consider the air quality data, which are provided as online sup-plementary material for Liu et al. (2022). For each of the 30 Chinese cities, a time series of daily air quality levels (December 2013--July 2019, so length = 2068) is reported. Here, air quality is measured on an ordinal scale with levels $D_t =0$ if the $t$-th daily air quality is `excellent' to $D_t =5$ if the $t$-th daily air quality is `severely polluted'.

We consider the count time series $(D_t)_{1\leq t\leq 2068}$ corresponding to Zhengzhou city as an example. Figure 4 (a) presents the plot of the daily air quality data $\{D_t\}_{t=1}^{2068}$, which exhibit strong positive autocorrelations and weak PACF values, see Figure 4 (b, c). The sample ACF and PACF imply that the MVJ$(p_1,p_2)$ models with $p_1\geq 1$ and $p_2\geq 1$ should be considered. Moreover, $D_t\in\{0,1,2,3,4,5\}$. Thus, $d=5$.

We use the first $T=1568$ observations for fitting the MVJ models with the clipped-Laplace link function $\mathrm{CL}_1(\cdot)$ and leave out the last $T_{new}=500$ observations for
a later forecast experiment. For model selection, we consider MVJ$(p_1,p_2)$ models with $(p_1,p_2)=(1,0)$, $(1,1)$, $(1,2)$, $(2,0)$, $(2,1)$, and $(2,2)$ as candidate models. The AIC and BIC values of the candidate models for the air quality counts are given in Table 6. It is easy to see that AIC select the
MVJ(1,2) model while BIC select the MVJ(1,0). However the BIC values of the MVJ(1,0) and  MVJ(1,2) models are very close. Based on these considerations, we adopt the MVJ(1,2) model to analyze the air quality data.

The final estimates together with their estimated standard deviations (SDs) are summarized in Table 7.  Obviously, the OWLS estimate $\breve{\vtheta}_T\notin \Theta_1$
and the OLS estimate $\hat{\vtheta}_T\notin \Theta_1$, which implies that the fitted MVJ(1,2) model has a nonlinear mean structure. Moreover, the OWLS estimator $\breve{\vtheta}_T$ and the OLS estimator $\hat{\vtheta}_T$ give similar SDs for $\vtheta$ . Figure 4(d) presents the standardized residuals plot using the OLS estimates of $\vtheta$ and $\vvartheta$.

To check the adequacy of the fitted models, we consider the approaches discussed in Section 3.5. The model diagnostics statistics MAR and MSPR with the OLS and OWLS estimates of $\vtheta$ are presented in Table 8. The sample means ($E_T(r_t)$), sample SDs ($Sd_T(r_t)$), and the maximum absolute value of the sample autocorrelation ($\max_{1\leq k\leq 20}|\rho_{r_t}(k)|$) of the standardized residuals using the OLS and OWLS estimates of $\vtheta$ are also reported. From Table 8, we see that, the OLS and OWLS methods have the same MAR values but the MSPR value of the OLS method is closer to 1 than that of the OWLS method using the MVJ(1,2) model, which indicates a better model fit.

Finally, let us analyze the forecast performance of the fitted MVJ(1,2) model with the OLS and OWLS estimation methods. We apply the MAR and MSPR criteria to the 500 new air quality counts. The sample means ($E_{T_{new}}(r_t)$), sample SDs ($Sd_{T_{new}}(r_t)$), and the maximum absolute value of the sample autocorrelation ($\max_{1\leq k\leq 16}|\rho_{r_t}(k)|$) of the standardized prediction residuals using the OLS and OWLS estimates of $\vtheta$ are also considered. Results are summarized in Table 9.  Obviously, the OWLS-fitted MVJ(1,2) model shows the better predictive performance regarding the 500 air quality counts, according to the MAR criterion.

\section{Discussion} \setcounter {equation}{0}
\def\theequation{\thesection.\arabic{equation}}

In this paper, we presented MVJ models tailored for the analysis of bounded integer-valued time series. These models offer flexible mean and variance structures, enhancing their applicability. Notably, our MVJ model incorporates a clipped-Laplace link, which simplifies to a linear mean model for positive parameter values summing to one. However, it also accommodates negative parameters, thus enabling the modeling of autocorrelation function (ACF) values.

We rigorously established conditions for stationarity and ergodicity, providing a robust theoretical framework. Furthermore, we demonstrated the consistency and asymptotic normality of the conditional least squares estimators, ensuring reliable parameter estimation.

To facilitate model selection, we proposed two criteria tailored to MVJ models. Additionally, we introduced two diagnostic statistics for evaluating model performance. Through empirical analyses of real data sets, we demonstrated the efficacy of MVJ models in generating more realistic forecasts for both mean and variance.

Our findings underscore the utility of MVJ models in capturing the complexities inherent in bounded integer-valued time series, thereby improving forecasting accuracy and enhancing analytical insights.

\section*{Supplementary Material}

The online Supplementary Material contains the proofs of all Theorems, the detailed computation algorithm for the OLS and OWLS estimates and their estimated asymptotic covariance matrices, and additional simulation results.

\section*{Acknowledgements}

We thank Professors Danning Li and Lianyan Fu for help discussions.


\makeatletter
\renewenvironment{thebibliography}[1]
{\section*{\refname}%
\@mkboth{\MakeUppercase\refname}{\MakeUppercase\refname}%
\list{\@biblabel{\@arabic\c@enumiv}}%
{\settowidth\labelwidth{\@biblabel{#1}}%
\leftmargin\labelwidth \advance\leftmargin\labelsep
\advance\leftmargin by 2em%
\itemindent -2em%
\@openbib@code
\usecounter{enumiv}%
\let\p@enumiv\@empty
\renewcommand\theenumiv{\@arabic\c@enumiv}}%
\sloppy \clubpenalty4000 \@clubpenalty \clubpenalty
\widowpenalty4000%
\sfcode`\.\@m} {\def\@noitemerr
{\@latex@warning{Empty `thebibliography' environment}}%
\endlist}
\renewcommand\@biblabel[1]{}
\makeatother

\begin{thebibliography}{}

\bibitem{Aknouche2024}
Aknouche, A. \& Scotto, M. G. (2024) A multiplicative thinning-based integer-valued GARCH model. {\it  Journal of Time Series Analysis}  \textbf{45}: 4--26.


\bibitem{Billingsley1968}
Billingsley, P. (1968) Convergence of probability measures. John Wiley, New York.

\bibitem{Cai2017}
Cai, Z., He, X., Sun, J. \& Vasconcelos, N. (2017)
Deep learning with low precision by half-
wave Gaussian quantization. In: 2017 IEEE
Conference on Computer Vision and Pattern Recognition (CVPR), Honolulu, HI, pp.
5406--5414.

\bibitem{Chen2020}
Chen, H., Li, Q. \& Zhu, F. (2020) Two classes of dynamic binomial integer-valued ARCH
models. {\it Brazilian Journal of Probability and Statistics} \textbf{34}: 685--711.


\bibitem{Chen2023}
Chen, H. (2023) A new soft-clipping discrete Beta GARCH model and its application on Measles infection. {\it Statistics }\textbf{6}: 293--311.

\bibitem{Jentsch2019}
Jentsch, C. \& Reichmann, L. (2019) Generalized binary time series models. {\it Econometrics} \textbf{7}: 47.

\bibitem{Kang2023}
Kang, Y., Wang, S., Wang, D. \& Zhu, F. (2023) Analysis of zero-and-one inflated bounded count time
series with applications to climate and crime data. {\it Test} \textbf{32}:34--73.

\bibitem{Kedem2002}
Kedem, B. \&   Fokianos, K. (2002) Regression Models for Time Series Analysis.  Hoboken, New Jersey: Wiley.

\bibitem{Klimek2020}
Klimek, M. D. \& Perelstein, M. (2020) Neural network-based approach to phase space integration. {\it SciPost Physics} \textbf{9}: 053.

\bibitem{Liu2022}
Liu, M., Zhu, F. \& Zhu, K. (2022) Modeling normalcy-dominant ordinal time series: An application to air quality level. {\it Journal of
Time Series Analysis} \textbf{43}: 460--478.

\bibitem{Liu2013}
Liu, T. \& Yuan, X. (2013) Random rounded integer-valued autoregressive conditional heteroskedastic process. {\it Statistical Papers} \textbf{54}: 645--683.


\bibitem{Liu2024}
Liu, T. \& Yuan, X. (2024) Semiparametric mean and variance joint models with Laplace link for count time series. arXiv:2404.18421v1 [stat.ME].


\bibitem{McKenzie1985}
McKenzie, E. (1985) Some simple models for discrete variate time series. {\it Journal of the American Water Resources Bulletin} \textbf{21}: 645--650.


\bibitem{Meyn2009}
Meyn, S. \& Tweedie, R. L. (2009) Markov Chains and Stochastic Stability, second ed.
Cambridge University Press, Cambridge.

\bibitem{Newey1994}
Newey, W. K. \& McFadden, D. (1994) Large sample estimation and hypothesis testing, in: R. Engle, D. McFadden (Eds.), Handbook of Econometrics, Vol. IV,
Elsevier, Amsterdam, pp. 2111-2245.

\bibitem{Risti2016}
Risti\'{c}, M. M., Wei${\ss}$, C.~H. \& Janji\'{c}, A. D. (2016) A binomial integer-valued ARCH model. {\it International Journal of Biostatistics} \textbf{12}: 20150051.

\bibitem{Wei2018}
Wei${\ss}$, C.~H. (2018) {\it An Introduction to Discrete-Valued Time Series.} John Wiley \& Sons, Chichester.


\bibitem{WeiJahn2022}
Wei${\ss}$, C. H. \& Jahn, M. (2022) Soft-clipping INGARCH models for time series of
bounded counts. {\it Statistical Modelling} https://doi.org/10.1177/1471082X221121223.

\bibitem{Zheng2015}
Zheng, T., Xiao, H. \& Chen, R. (2015) Generalized ARMA models with martingale difference errors. {\it Journal of Econometrics} \textbf{189}: 492--506.



\end{thebibliography}


\newpage
\begin{table*}\begin{center}
Table 1: ACF values for the simulated data.\\
\label{tab:1}
\footnotesize
\begin{tabular}{*{9}{c}}
\hline \multicolumn{1}{c}{Setting}&\multicolumn{1}{c}{$n$}&\multicolumn{1}{c}{}&
\multicolumn{6}{l}{MVJ$(p_1,p_2)$ }\\
\cmidrule(l){4-9}
   &  && $M_1$  &  $M_2$     & $M_3$  & $M_4$  &$M_5$  &$M_6$ \\
 \hline
 (a)   &500&$\rho_{X_t}(1)$  &0.572 &0.337  &0.328   &0.346 &0.145 &0.329   \\
               &&$\rho_{X_t}(2)$  &0.321 &0.214  &0.189   &0.534 &0.422 &0.479  \\\\
 (b)   &500&$\rho_{X_t}(1)$  &-0.429   &-0.348  &-0.260   &-0.136   &-0.080    &-0.092   \\
             &&$\rho_{X_t}(2)$  &0.091   &0.152  &0.071   &-0.339   &-0.295    &-0.334       \\
                 \hline
\end{tabular}\end{center}
\end{table*}

\begin{table*}\begin{center}
Table 2: Mean  of estimates, RMSE (within parentheses) for the MVJ models with $\vartheta_1=1/2$ and $\vartheta_2=1/3$ under setting (a).\\
\label{tab:1}
\footnotesize
\begin{tabular}{*{10}{c}}
\hline
$M(p_1,p_2)$& T & Method &$c$            & $\phi_1$     & $\phi_2$ & $\psi_1$ & $\psi_2$ & $\vartheta_1$ & $\vartheta_2$\\
\hline
$M_1$(1,0)    &   &  True Value     &-0.2     &0.5   &          & &  &1/2 &1/3\\
     &200&  OLS  &-0.2307  &0.4698  &          & && 0.4865  & 0.2916 \\
     &   &       &(0.2802)&(0.1309)&      & & &(0.1533) &(0.1595)\\
     &   &  OWLS  &-0.2318  &0.4687  &          && & &\\
     &   &       &(0.2786)&(0.1323)&      && & &\\
     &500&  OLS  &-0.2184  &0.4907  &          & && 0.4932 & 0.3231 \\
     & &         &(0.1628)&(0.0891)&       & && (0.0780)&(0.0907)\\
     &   &  OWLS  &-0.2187  &0.4907  &          && & &\\
     &   &       &(0.1618)&(0.0883)&      && & &\\\\
$M_2$(1,1)   & &   True Value  &-0.2&0.4&          &0.4 & & 1/2 &1/3 \\
     &200&  OLS  &-0.1072&0.4005  &          &0.3299 &&  0.4554 &0.3328 \\
     &   &       &(0.3032)&(0.1122)&      &(0.2392)& &(0.2460) &(0.1093)\\
     &   &  OWLS  &-0.1100   &0.4005  &          &0.3279& & &\\
     &   &       &(0.3023)&(0.1126)&      &(0.2370)&& &\\
     &500&  OLS  &-0.1618  &0.3976  &          &0.3751 && 0.4763& 0.3372 \\
     & &         &(0.1781)&(0.0662)&       &(0.1314) && (0.1253)&(0.0615)\\
     &   &  OWLS  &-0.1628  &0.3981  &          &0.3743& & &\\
     &   &       &(0.1771)&(0.0658)&      &(0.1308)& & &\\\\
$M_3$(1,2)   & &  True Value     &-0.2 &0.4&     &0.1 &0.4&1/2&1/3 \\
     &200&  OLS  &0.0292  &0.3931  & &0.1318 &0.2710 & 0.4431 & 0.3395 \\
     &   &       &(0.4810)&(0.1069)&      &(0.3766) & (0.3328)&(0.3123) &(0.0939)\\
     &   &  OWLS  &0.0259   &0.3928  &  &0.1317&0.2698 & &\\
     &   &       &(0.4808)&(0.1055)&      &(0.3743)&(0.3302) & &\\
     &500&  OLS  &-0.1283  &0.3944  &          &0.1040 &0.3711&  0.4648& 0.3382 \\
     & &         &(0.2080)&(0.0605)&       &(0.1737) &(0.1570)& (0.1971)&(0.0614)\\
     &   &  OWLS  &-0.1292  &0.3948  &          &0.1038&0.3700 & &\\
     &   &       &(0.2072)&(0.0599)&      &(0.1727)&(0.1567) & &\\
  \hline
\end{tabular}\end{center}
\end{table*}

\begin{table*}\begin{center}
Table 2 (continued): Mean  of estimates, RMSE (within parentheses) for the MVJ models with $\vartheta_1=1/2$ and $\vartheta_2=1/3$ under setting (a).\\
\label{tab:1c}
\footnotesize
\begin{tabular}{*{10}{c}}
\hline
$M(p_1,p_2)$& T & Method &$c$            & $\phi_1$     & $\phi_2$ & $\psi_1$ & $\psi_2$ &$\vartheta_1$ & $\vartheta_2$\\
\hline
$M_4$(2,0)   & & True Value   &-0.2&0.2&0.5    & & &1/2&1/3\\
     &200&  OLS  &-0.1973  &0.1826  &0.4603          & && 0.4488  & 0.3221 \\
     &   &       &(0.2829)&(0.1389)&(0.1298)      & & &(0.1659) &(0.1260)\\
     &   &  OWLS  &-0.1998   &0.1838  &0.4598          && & &\\
     &   &       &(0.2810)&(0.1360)& (0.1297)     && & &\\
     &500&  OLS  &-0.1969  &0.1936  &0.4841          & && 0.4774 & 0.3315 \\
     & &         &(0.1775)&(0.0642)&(0.0776)       & && (0.0983)&(0.0699)\\
     &   &  OWLS  &-0.1976  &0.1943  &0.4841          && & &\\
     &   &       &(0.1761)&(0.0624)&(0.0768)      && & &\\\\
$M_5$(2,1)   & & True Value   &-0.2&0.1&0.4    &0.4& &1/2&1/3\\
     &200&  OLS  &-0.0850  &0.0945  &0.3850     &0.3598 &&  0.4647 & 0.3314  \\
     &   &       &(0.3503)&(0.1020)&(0.1136)      &(0.2098) & &(0.2841) &(0.0922)\\
     &   &  OWLS  &-0.0896   &0.0969  &0.3847          &0.3564& & &\\
     &   &       &(0.3509)&(0.0958)& (0.1127)     &(0.2089)& & &\\
     &500&  OLS  &-0.1572  &0.1003  &0.3943          &0.3849 &&   0.4730&0.3372  \\
     & &         &(0.1851)&(0.0578)& (0.0718)      &(0.1105) && (0.1511)&(0.0513)\\
     &   &  OWLS  &-0.1584  &0.1010  &0.3945          &0.3836& & &\\
     &   &       &(0.1841)&(0.0560)&(0.0712)      &(0.1101)& & &\\\\
$M_6$(2,2)   & & True Value   &-0.2&0.1&0.4    &0.1&0.3&1/2&1/3\\
     &200&  OLS  &0.0110  &0.0942  &0.3910          & 0.1375 &0.1825&0.4507 &0.3351 \\
     &   &       &(0.4991)&(0.1026)&   (0.1130)   &(0.3418) &(0.3099) &(0.2855) &(0.0889)\\
     &   &  OWLS  &0.0103   &0.0978  &0.3890          &0.1355&0.1804 & &\\
     &   &       &(0.5139)&(0.1068)&   (0.1125)   &(0.3457)&(0.3108) & &\\
     &500&  OLS  &-0.1205  &0.0992  &0.3979       &0.1211 &0.2477& 0.4651 & 0.3388\\
     & &         &(0.2225)&(0.0551)&   (0.0646)    & (0.1761)&(0.1613)& (0.1636)&(0.0520)\\
     &   &  OWLS  &-0.1212  &0.0999  &0.3981          &0.1205&0.2468 & &\\
     &   &       &(0.2213)&(0.0533)&  (0.0642)  &(0.1755)&(0.1609) & &\\\\
  \hline
\end{tabular}\end{center}
\end{table*}

\begin{table*}\begin{center}
Table 3: Mean  of estimates, RMSE (within parentheses) for the MVJ models with $\vartheta_1=1/2$ and $\vartheta_2=1/3$ under setting (b).\\
\label{tab:3}
\footnotesize
\begin{tabular}{*{10}{c}}
\hline
$M(p_1,p_2)$& T & Method &$c$            & $\phi_1$     & $\phi_2$ & $\psi_1$ & $\psi_2$ & $\vartheta_1$ & $\vartheta_2$\\
\hline
$M_1$(1,0)    &   &  True Value     &5     &-0.5   &          & &  &1/2 &1/3\\
     &200&  OLS  &5.0626& -0.5270 &          & && 0.3685 & 0.3549 \\
     & &         &(0.4759)&(0.1239)&       & && (0.4923)&(0.0959)\\
     &   &  OWLS  &5.0619  &-0.5247  &          && & &\\
     &   &       &(0.4770)&(0.1055)&      && & &\\
     &500&  OLS  &5.0107  &-0.5095  &          & &&0.4072  & 0.3502 \\
     &   &       &(0.2888)&(0.0688)&      & & &(0.3121) &(0.0616)\\
     &   &  OWLS  &5.0120   &-0.5087  &          && & &\\
     &   &       &(0.2920)&(0.0595)&      && & &\\ \\
$M_2$(1,1)   & &   True Value  &5&-0.4&          &-0.4 & & 1/2 &1/3 \\
     &200&  OLS  &4.8036  &-0.4635  &          &-0.3107 &&  0.3777 & 0.3543\\
     &   &       &(0.8664)&(0.1966)&      &(0.2862)& &(0.4339) &(0.0989)\\
     &   &  OWLS  &4.8003   &-0.4522  &          &-0.3172& & &\\
     &   &       &(0.8826)&(0.1759)&      &(0.3480)&& &\\
     &500&  OLS  &4.9390  &-0.4229  & &-0.3726 && 0.4086 & 0.3499 \\
     & &         &(0.4322)&(0.0954)&       &(0.1399) && (0.2778)&(0.0615)\\
     &   &  OWLS  &4.9392  &-0.4180  &  &-0.3719& & &\\
     &   &       &(0.4323)&(0.0797)&      &(0.1405)& & &\\\\
$M_3$(1,2)   & &  True Value     &5 &-0.4&     &-0.1 &-0.4&1/2&1/3\\
     &200&  OLS  &4.1919  &-0.4718  &          &0.0407 &-0.2384& 0.3479 & 0.3711 \\
     & &         &(1.7257)&(0.2035)&       &(0.3077) &(0.3763)& (0.6234)&(0.1232)\\
     &   &  OWLS  &4.2204  &-0.4655  &          &0.0421&-0.2420 & &\\
     &   &       &(1.7696)&(0.1994)&      &(0.3196)&(0.3756) & &\\
     &500&  OLS  &4.6110  &-0.4309  &          &-0.0284 &-0.3331& 0.3823  & 0.3612 \\
     &   &       &(0.8668)&(0.1037)&      &(0.1628) & (0.1813)&(0.3118) &(0.0761)\\
     &   &  OWLS  &4.6091   &-0.4239  &        &-0.0282&-0.3355 & &\\
     &   &       &(0.8700)&(0.0867)&      &(0.1655)&(0.1819) & &\\
  \hline
\end{tabular}\end{center}
\end{table*}

\begin{table*}\begin{center}
Table 3 (continued): Mean  of estimates, RMSE (within parentheses) for the MVJ models with $\vartheta_1=1/2$ and $\vartheta_2=1/3$ under setting (b).\\
\label{tab:3c}
\footnotesize
\begin{tabular}{*{10}{c}}
\hline
$M(p_1,p_2)$& T & Method &$c$            & $\phi_1$     & $\phi_2$ & $\psi_1$ & $\psi_2$ & $\vartheta_1$ & $\vartheta_2$\\
\hline
$M_4$(2,0)   & & True Value   &5&-0.2&-0.5    & & &1/2&1/3\\
     &200&  OLS  &5.0594  &-0.1986  &-0.5315          & && 0.3565  & 0.3613 \\
     &   &       &(0.5650)&(0.0797)&(0.1353)      & & &(0.4500) &(0.0967)\\
     &   &  OWLS  &5.0661   &-0.2049  &-0.5299          && & &\\
     &   &       &(0.5964)&(0.1092)& (0.1185)     && & &\\
     &500&  OLS  &5.0362  &-0.2037  &-0.5143      & && 0.3791 & 0.3566 \\
     & &         &(0.3429)&(0.0474)&(0.0796)       & && (0.3008)&(0.0638)\\
     &   &  OWLS  &5.0361  &-0.2038  &-0.5128      && & &\\
     &   &       &(0.3431)&(0.0481)&(0.0693)      && & &\\\\
$M_5$(2,1)   & & True Value   &5&-0.1&-0.4    &-0.4& &1/2&1/3\\
     &200&  OLS  &4.6698  &-0.1084  &-0.4231          &-0.2734 && 0.3566  & 0.3670 \\
     &   &       &(1.2320)&(0.0955)&(0.1436)      &(0.3698) & &(0.4541) &(0.1102)\\
     &   &  OWLS  &4.6703   &-0.1096  &-0.4175         &-0.2769& & &\\
     &   &       &(1.2339)&(0.1033)& (0.1237)     &(0.3664)& & &\\
     &500&  OLS  &4.8783  &-0.0990  & -0.4106          &-0.3627 && 0.3860 & 0.3589 \\
     & &         &(0.6196)&(0.0504)&(0.0736)       & (0.1755)&& (0.2876)&(0.0713)\\
     &   &  OWLS  &4.8784  &-0.0992  &-0.4084          &-0.3635& & &\\
     &   &       &(0.6198)&(0.0589)&(0.0660)      &(0.1767)&& &\\\\
$M_6$(2,2)   & & True Value   &5&-0.1&-0.4    &-0.1&-0.3&1/2&1/3\\
     &200&  OLS  &4.6402  &-0.1115  &-0.4735          &-0.0372 &-0.2125& 0.3080  &0.3761  \\
     &   &       &(1.2735)&(0.0957)&(0.2337)      &(0.2818) &(0.3657) &(0.4601) &(0.1082)\\
     &   &  OWLS  &4.6417   &-0.1185  &-0.4950         &-0.0338&-0.2344 & &\\
     &   &       &(1.2737)&(0.2213)& (0.7802)     &(0.3231)&(0.7242) & &\\
     &500&  OLS  &4.8634  &-0.1057  &-0.4269          &-0.0769 &-0.2654& 0.3897 & 0.3584 \\
     & &         &(0.6294)&(0.0525)&(0.1006)       &(0.1433) &(0.1684)& (0.2754)&(0.0658)\\
     &   &  OWLS  &4.8647  &-0.1068  &-0.4233          &-0.0751&-0.2645 & &\\
     &   &       &(0.6289)&(0.0626)&(0.0892)      &(0.1582)&(0.1730) & &\\\\
  \hline
\end{tabular}\end{center}
\end{table*}

\begin{table*}\begin{center}
Table 4 : Frequency of orders selected by AIC and BIC for the MVJ models with $\vartheta_1=1/2$ and $\vartheta_2=1/3$ under setting (a) in 1000 realizations.\\
\label{tab:4}
\scriptsize
\begin{tabular}{*{9}{c}}
\hline \multicolumn{1}{c}{Model}&\multicolumn{1}{c}{ T
}&\multicolumn{1}{c}{ }&
\multicolumn{5}{l}{$(p_1,p_2)$ }\\
\cmidrule(l){4-9}
 &  &   & (1,0)  &  (1,1)     & (1,2)  & (2,0)  &(2,1)  &(2,2) \\
 \hline $M_1$   &200 & AIC  &390  &145   &164   &93   &61   &147   \\
        &    & BIC  &676  &80   &85   &90   &29  &40  \\
        &500 & AIC  &481  &167  &127  &67   &83   &75   \\
        &    & BIC  &823  &47   &29   &72  &21   &8   \\
        &800 & AIC  &464   &210  &134  &69   &53   &70   \\
        &    & BIC  &853   &58  &15  &57  &8   &9   \\
        \\
\hline $M_2$   &200 & AIC  &214  &311   &161   &145   &58   &111   \\
        &    & BIC  &503  &224   &54   &188   &12  &19\\
        &500 & AIC  &37  &495  &129  &200   &64   &75   \\
        &    & BIC  &171  &520  &25  &268  &10   &6   \\
        &800 & AIC  &5   &561  &123  &172   &51   &88   \\
        &    & BIC  &53  &664  &29  &236  &11   &7   \\
        \\
  \hline $M_3$ &200 & AIC  &148  &280   &369   &34   &64   &105   \\
        &    & BIC  &439  &337   &145   &42   &17  &20  \\
        &500 & AIC  &9  &153  &619  &1   &65   &153   \\
        &    & BIC  &113 &384  &454 &5  &31   &13   \\
        &800 & AIC  &1  &44  &693  &0   &70   &192   \\
        &    & BIC  &12 &233  &686  &3  &56   &10   \\
        \\
 \hline $M_4$&200 & AIC  &3  &18   &71   &485   &152   &271   \\
        &    & BIC  &17  &26 &54   &759  &54  &90  \\
        &500 & AIC  &0  &1  &4  &519   &197   &279   \\
        &    & BIC  &0  &2   &2   &905  &50   &41   \\
        &800 & AIC  &0   &1  &4  &548   &187   &260   \\
        &    & BIC  &0   &2  &6  &921  &42   &29   \\
        \\
  \hline $M_5$    &200 & AIC  &0  &27   &73   &236   &440   &224   \\
        &    & BIC  &0  &66   &63   &520   &299  &52  \\
        &500 & AIC  &0  &1  &11  &23   &688   &277   \\
        &    & BIC  &0  &5   &18  &136  &804   &37   \\
        &800 & AIC  &0   &0  &0  &1   &745   &254   \\
        &    & BIC  &0   &0  &1  &31  &922   &46   \\
        \\
  \hline $M_6$ &200 & AIC  &1  &36   &52   &302   &290   &319   \\
        &    & BIC  &4  &83   &39   &628   &186  &60  \\
        &500 & AIC  &0  &1  &3  &56   &260   &680   \\
        &    & BIC  &0  &8   &4   &324  &407   &257   \\
        &800 & AIC  &0   &0  &1  &5   &112   &882   \\
        &    & BIC  &0   &0  &1  &128  &368   &503   \\
        \hline
\end{tabular}\end{center}
\end{table*}

\begin{table*}\begin{center}
Table 5 : Frequency of orders selected by AIC and BIC for the MVJ models with $\vartheta_1=1/2$ and $\vartheta_2=1/3$ under setting (b) in 1000 realizations.\\
\label{tab:5}
\scriptsize
\begin{tabular}{*{9}{c}}
\hline \multicolumn{1}{c}{Model}&\multicolumn{1}{c}{ T
}&\multicolumn{1}{c}{ }&
\multicolumn{5}{l}{$(p_1,p_2)$ }\\
\cmidrule(l){4-9}
 &  &   & (1,0)  &  (1,1)     & (1,2)  & (2,0)  &(2,1)  &(2,2) \\
 \hline $M_1$&200 & AIC  &523  &106   &94   &133   &85   &59   \\
        &    & BIC  &863  &35   &8   &67   &16  &11  \\
        &500 & AIC  &597  &113  &73  &126   &56   &35   \\
        &    & BIC  &938  &12   &5   &40  &5   &0   \\
        &800 & AIC  &576   &114  &80  &152   &60   &18   \\
        &    & BIC  &934   &15  &4  &42  &5   &0   \\
        \\
\hline $M_2$&200 & AIC  &291  &273   &122   &193   &52   &69   \\
        &    & BIC  &625  &169   &33   &163   &5  &5  \\
        &500 & AIC  &79  &497  &118  &222   &29   &55   \\
        &    & BIC  &343  &423   &10   &222  &1   &1   \\
        &800 & AIC  &18   &628  &110  &178   &11   &55   \\
        &    & BIC  &164   &625  &8  &202  &0   &1   \\
        \\
  \hline $M_3$&200 & AIC  &426  &45   &259   &91   &55   &124   \\
        &    & BIC  &823  &11   &70   &59   &17  &20  \\
        &500 & AIC  &229  &15  &571  &39   &35   &111   \\
        &    & BIC  &731  &1   &204   &47  &8   &9   \\
        &800 & AIC  &126   &5  &750  &14   &14   &91   \\
        &    & BIC  &597   &0  &365  &30  &1   &7   \\
        \\
 \hline $M_4$&200 & AIC  &0  &1   &15   &669   &169   &146   \\
        &    & BIC  &0  &7   &11   &934   &39  &9  \\
        &500 & AIC  &0  &0  &0 &689   &174   &137   \\
        &    & BIC  &0  &0  &0 &964  &33   & 3  \\
        &800 & AIC  &0   &0  &0  &705   &191   &104   \\
        &    & BIC  &0   &0  &0  &977  &23   &0   \\
        \\
  \hline $M_5$ &200 & AIC  &0  &13   &11   &363   &421   &192   \\
        &    & BIC  &3  &21   &8   &726   &218  &24  \\
        &500 & AIC  &0  &0  &1  & 154  & 686  & 159  \\
        &    & BIC  &0  &0   &1   &487  &496   &16   \\
        &800 & AIC  &0   &0  &0  &51   &796   &153   \\
        &    & BIC  &0   &0  &0  &297  &693   &10   \\
        \\
  \hline $M_6$&200 & AIC  &0  &3   &10   &561   &127   &299   \\
        &    & BIC  & 1 &6   &9   &885   &37  &62  \\
        &500 & AIC  &0  &0  &0  &325   &91   &584   \\
        &    & BIC  &0  &0  &0  &869  &27   &104   \\
        &800 & AIC  &0   &0  &0  &173   &56   &771   \\
        &    & BIC  &0   &0  &0 &739  &24   &237   \\
        \hline
\end{tabular}\end{center}
\end{table*}

\begin{table*}\begin{center}
Table 6: AIC and BIC values for the real data.\\
\label{tab:6}
\footnotesize
\begin{tabular}{*{9}{c}}
\hline \multicolumn{1}{c}{Data}&\multicolumn{1}{c}{$T$}&\multicolumn{1}{c}{}&
\multicolumn{6}{c}{MVJ$(p_1,p_2)$ }\\
\cmidrule(l){4-9}
   &  && (1,0)  &  (1,1)     & (1,2)  & (2,0)  &(2,1)  &(2,2) \\
 \hline
Geyser eruption  &249&AIC  &0.2352 &20.6654  &12.4636   &\textbf{-10.5217} &-6.3304 &-4.0682   \\
               &&BIC  &14.2566 &38.1920  &33.4711   &\textbf{6.9846} &14.6772 &20.4406  \\\\
Air quality  &1568&AIC  &149.4488   &150.2508  &\textbf{140.8118}   &151.3355   &154.7811    &145.2889   \\
             &&BIC  &\textbf{170.8714}   &177.0290  &172.9418   &178.1105   &186.9111    &182.7739      \\
                 \hline
\end{tabular}\end{center}
\end{table*}

\begin{table*}\begin{center}
Table 7: Estimates and their estimated standard deviations (in parentheses) for the real data.\\
\label{tab:7}
\scriptsize
\begin{tabular}{*{11}{c}}
\hline
Data& $T$ &Model&Estimate&$c$ & $\phi_1$ &$\phi_2$ & $\psi_1$ &$\psi_2$& $\vartheta_1$ & $\vartheta_2$\\
\hline
Geyser&  249  &(2,0) & OLS & 2.9132 &-0.4202 &0.4966 &&  &0.0849  &0.2328\\
            &    & &  &(0.4712) &(0.0750) &(0.0960) & &&   (0.0248)&(0.0273)\\
            &    &     & OWLS &2.9237 &-0.4187 &0.4960 & &&   &\\
            &    & &  &(0.4462) &(0.0695) &(0.0935) & &&   &\\
Air &  1568  &(1,2)  &OLS &0.2660 &0.6569 & &-0.0576  &0.2102&0.1961&0.1519\\
          &    &  & &(0.1256) &(0.0406) & &(0.0756) &(0.0711) &(0.0550)&(0.0347)\\
          &    &   &OWLS &0.2635 &0.6564 & &-0.0578  &0.2092  &&\\
          &    &  & &(0.1277) &(0.0405) & &(0.0754) &(0.0712) &  &\\
  \hline
\end{tabular}\end{center}
\end{table*}

\begin{table*}\begin{center}
Table 8:  Model diagnostics of the real data. Sample mean of the standardized residuals: $E_T(r_t)$; Sample SD of the standardized residuals: $Sd_T(r_t)$; the maximum absolute value of the sample autocorrelation: $\max_{1\leq k\leq 20}|\rho_{r_t}(k)|$.\\
\label{tab:8}
\footnotesize
\begin{tabular}{*{9}{c}}
\hline
Data& $T$ &Model&Method&$E_T(r_t)$&$Sd_T(r_t)$ &$\max_{1\leq k\leq 20}|\rho_{r_t}(k)|$& MAR & MSPR\\
\hline
Geyser eruption&  249  &(2,0) &OLS&0.0273  &0.9291 &0.136 &0.7234 &0.8605  \\
            &    &         &OWLS&0.0164  &0.9287 &0.136 &0.7234 &0.8593  \\
Air quality  &  1568 &(1,2)&OLS&-0.0064 &0.9997 &0.064 &0.8049 &0.9988 \\
            &    &         &OWLS&-0.0013 &0.9994 &0.064 &0.8049 &0.9981  \\
  \hline
\end{tabular}\end{center}
\end{table*}

\begin{table*}\begin{center}
Table 9: Predictions of the real data.\\
\label{tab:9}
\footnotesize
\begin{tabular}{*{9}{c}}
\hline
Data& $T_{new}$ &Model&Method&$E_{T_{new}}(r_t)$&$Sd_{T_{new}}(r_t)$ &$\max_{1\leq k\leq 16}|\rho_{r_t}(k)|$& MAR & MSPR\\
\hline
Geyser eruption&  50  &(2,0) &OLS&0.0870 &0.9989   &0.272&0.8044 &0.9856  \\
            &      &      &OWLS &0.0756   &0.9988  &0.271&0.8040 &0.9834  \\
Air quality  &  500  &(1,2)&OLS&-0.1111    &0.9815  &0.149 &0.7452 &0.9738  \\
            &    &        &OWLS&-0.1062   &0.9812  &0.149 &0.7441 &0.9721  \\
  \hline
\end{tabular}\end{center}
\end{table*}

\begin{center}
\begin{figure}
\centerline{\includegraphics[width=0.8\textwidth, angle=
0]{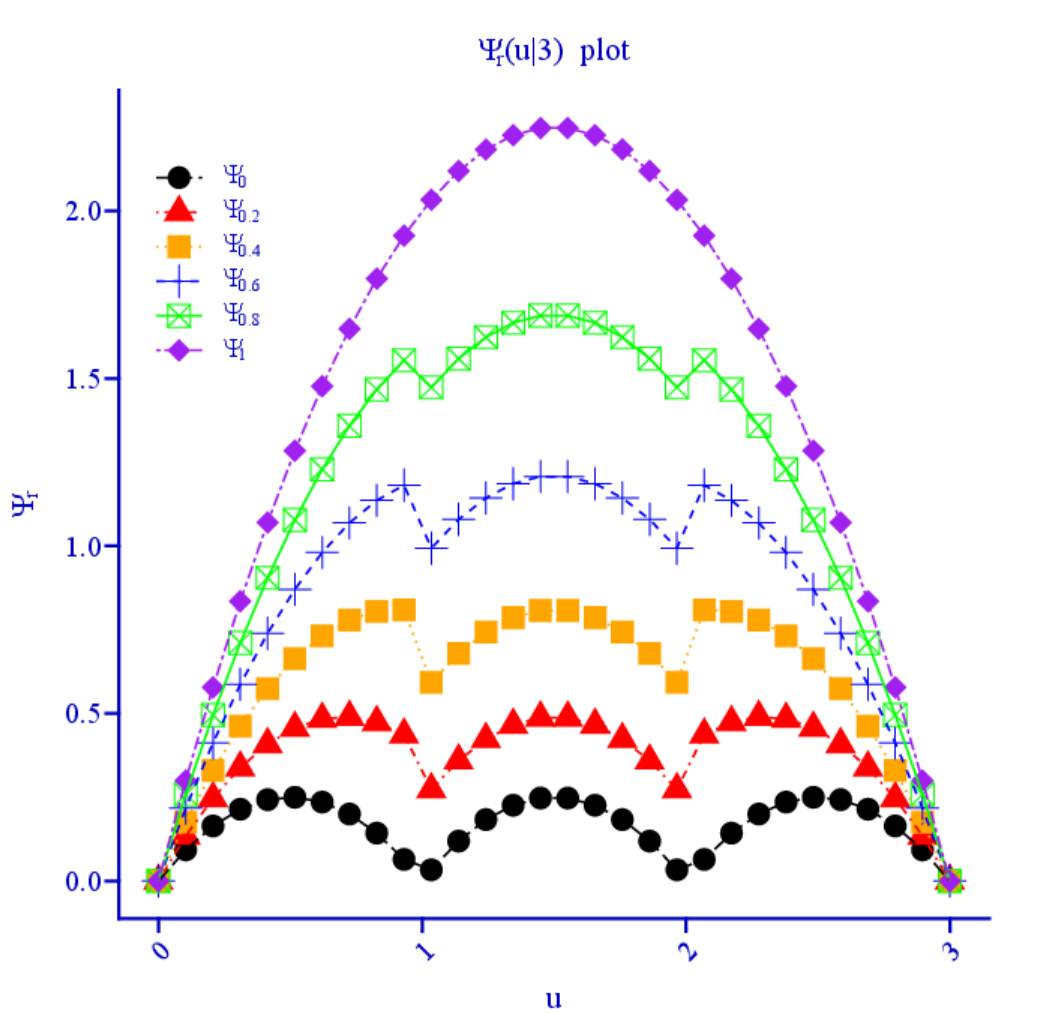}} \caption{$\Psi_r(u|3)$ plots with $u\in[0,3)$ and $r=0.0,0.2,0.4,0.6,0.8$ and $1.0$.}\label{Psiplot}
\end{figure}
\end{center}

\begin{center}
\begin{figure}
\centerline{\includegraphics[width=0.8\textwidth, angle=
0]{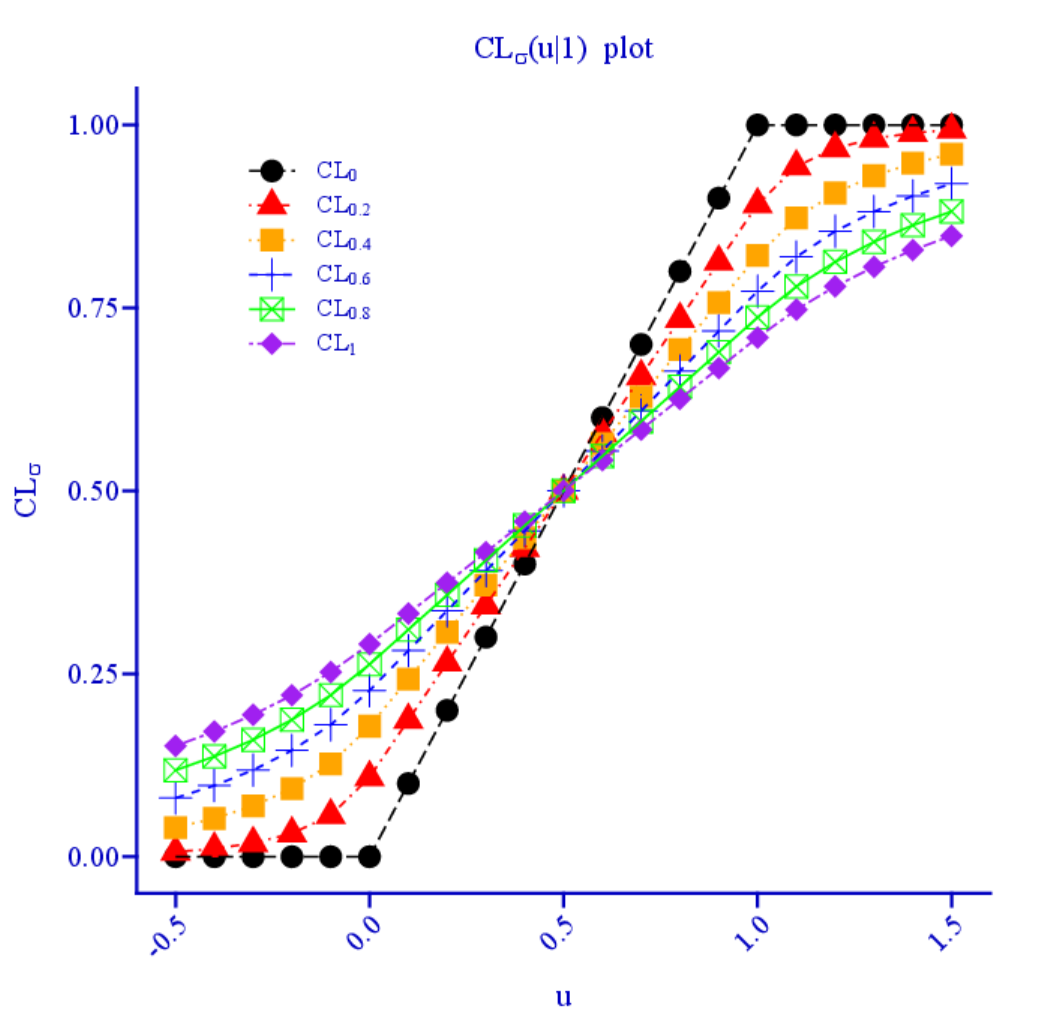}} \caption{$\mathrm{CL}_\sigma(u|1)$ plots with $u\in[-0.5,1.5]$ and $\sigma=0.0,0.2,0.4,0.6,0.8$ and $1.0$.}\label{CLplot}
\end{figure}
\end{center}

\begin{center}
\begin{figure}
\centerline{\includegraphics[width=0.8\textwidth, angle=
0]{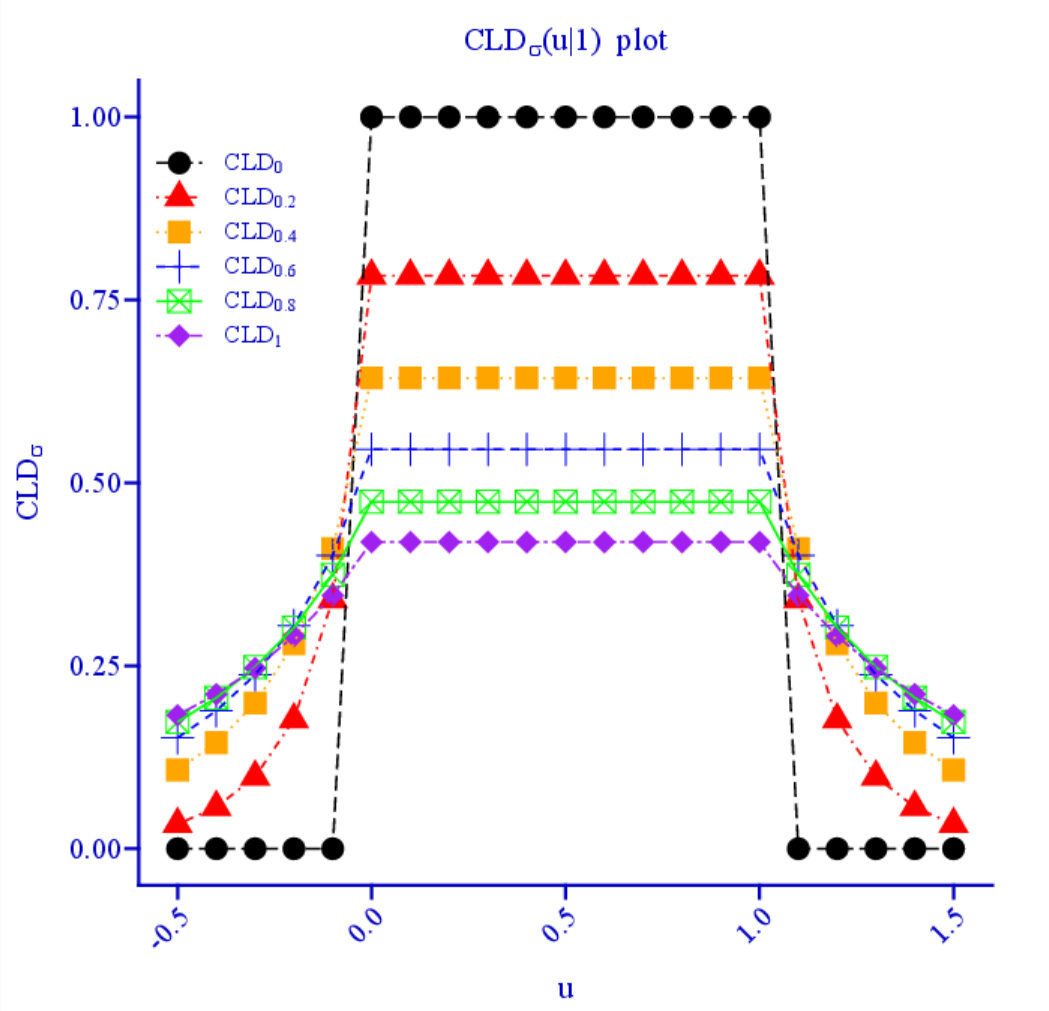}} \caption{$\mathrm{CLD}_\sigma(u|1)$ plots with $u\in[-0.5,1.5]$ and $\sigma=0.0,0.2,0.4,0.6,0.8$ and $1.0$.}\label{CLplot}
\end{figure}
\end{center}

\begin{center}
\begin{figure}
\centerline{\includegraphics[width=1.0\textwidth, angle=
0]{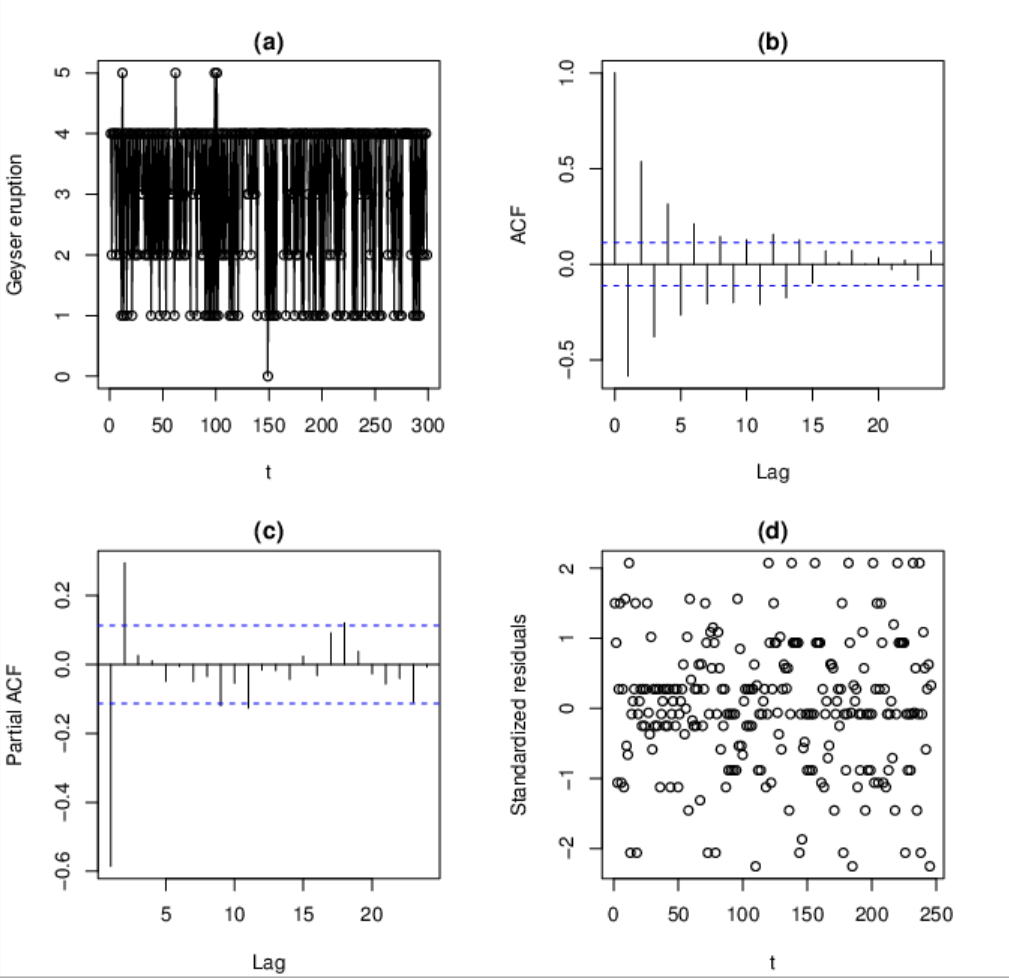}} \caption{Geyser eruption counts from Section 5.1: (a) time series plot; (b) sample ACF against Lag; (c) sample PACF against Lag; and (d) standardized residuals plot using the OLS estimates of $\vtheta$ and $\vvartheta$.}\label{Geyser}
\end{figure}
\end{center}

\begin{center}
\begin{figure}
\centerline{\includegraphics[width=1.0\textwidth, angle=
0]{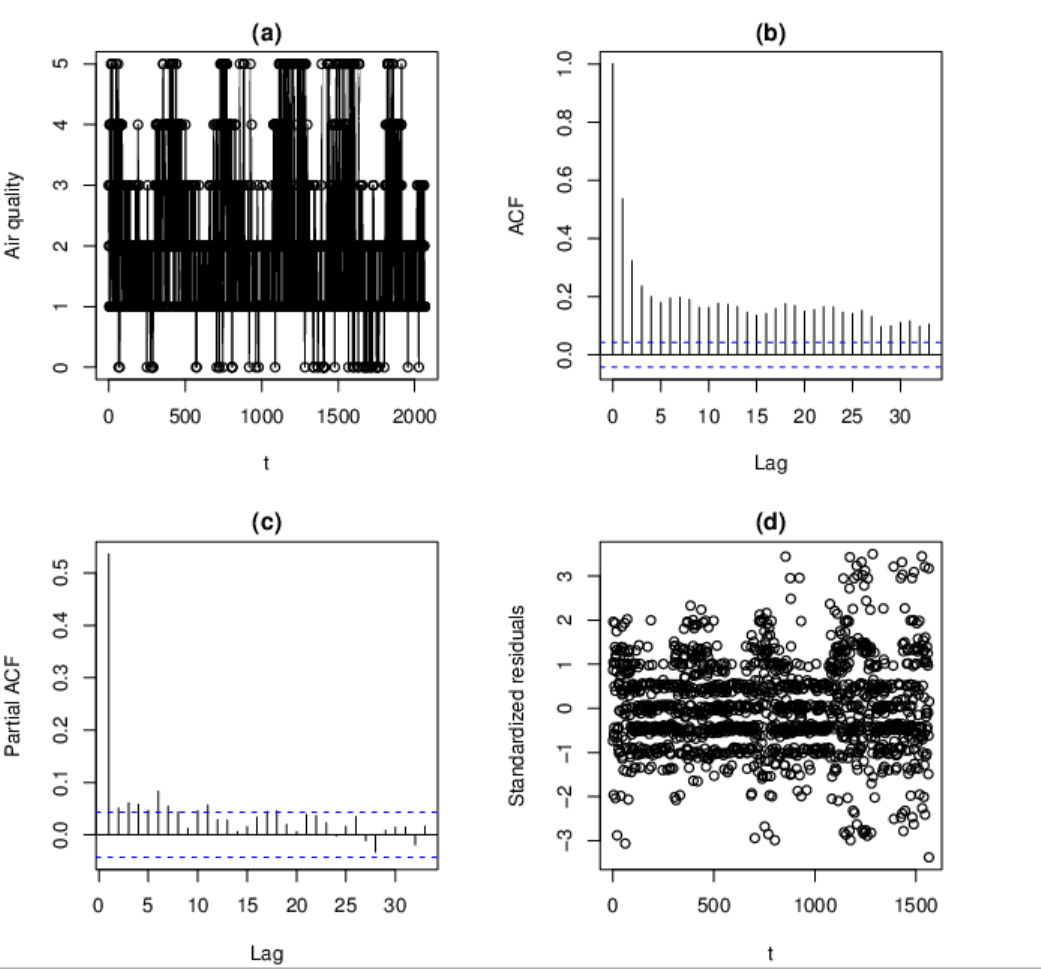}} \caption{Air quality counts from Section 5.2: (a) time series plot; (b) sample ACF against Lag; (c) sample PACF against Lag; and (d) standardized residuals plot using the OLS estimates of $\vtheta$ and $\vvartheta$.}\label{Air}
\end{figure}
\end{center}

\end{document}